\documentclass[review]{elsarticle}

\usepackage{color}
\usepackage[table,xcdraw]{xcolor}
\usepackage{adjustbox}
\usepackage{amsmath}
\usepackage{amssymb}
\begin{document}

\begin{frontmatter}

\title{Database of novel magnetic materials for high-performance permanent magnet development}


\author[ICCRAM]{P. Nieves}
\cortext[mycorrespondingauthor]{Corresponding author}
\ead{pnieves@ubu.es}

\address[ICCRAM]{ICCRAM, International Research Center in Critical Raw Materials and Advanced Industrial Technologies,
Universidad de Burgos, 09001 Burgos, Spain}

\author[ICCRAM,VSB]{S. Arapan}

\address[VSB]{VSB Tech Univ Ostrava, IT4Innovations, 17 Listopadu 15, CZ-70833 Ostrava, Czech Republic}

\author[UBU]{J. Maudes}
\author[UBU]{R. Marticorena}
\author[ICCRAM]{N. L. Del Br\'{i}o}

\address[UBU]{Department of Civil Engineering, Universidad de Burgos, 09006 Burgos, Spain}

\author[DUK]{A. Kovacs}

\address[DUK]{Center for Integrated Sensor Systems, Danube University Krems, Lower Austria, 3500 Krems, Austria}

\author[BCM]{C. Echevarria-Bonet}
\author[BCM]{D. Salazar}

\address[BCM]{BCMaterials, Basque Centre for Materials, Applications and Nanostructures, UPV/EHU Science Park, 48940 Leioa, Spain}

\author[TUDA]{J. Weischenberg}
\author[TUDA]{H. Zhang}

\address[TUDA]{Department of Materials and Geosciences, TU Darmstadt, Darmstadt, 64287, Germany}

\author[UPP1]{O. Yu. Vekilova}

\address[UPP1]{Department of Physics and Astronomy, Uppsala University, Box 516, 75121 Uppsala, Sweden}

\author[ICCRAM]{R. Serrano-L\'{o}pez}

\author[BCM]{J.M. Barandiaran}
\author[TUDA]{K. Skokov}
\author[TUDA]{O. Gutfleisch}

\author[UPP1,UPP2]{O. Eriksson}

\address[UPP2]{School of Science and Technology, \"{O}rebro University, SE-70182 \"{O}rebro, Sweden}

\author[UPP1]{H.C. Herper}

\author[DUK]{T. Schrefl}

\author[ICCRAM,ICAMCyL]{S. Cuesta-L\'{o}pez}

\address[ICAMCyL]{ICAMCyL Foundation International Center for Advanced Materials and Raw Materials of Castilla y Le\'{o}n, 24492 Cubillos del Sil, Le\'{o}n, Spain}

\begin{abstract}
This paper describes the open Novamag database that has been developed for the design of novel Rare-Earth free/lean permanent magnets. The database software technologies, its friendly graphical user interface, advanced search tools and available data are explained in detail. Following the philosophy and standards of Materials Genome Initiative, it contains significant results of novel magnetic phases with high magnetocrystalline anisotropy obtained by three computational high-throughput screening approaches based on a crystal structure prediction method using an Adaptive Genetic Algorithm, tetragonally distortion of cubic phases and tuning known phases by doping. Additionally, it also includes theoretical and experimental data about fundamental magnetic material properties such as magnetic moments, magnetocrystalline anisotropy energy, exchange parameters, Curie temperature, domain wall width, exchange stiffness, coercivity and maximum energy product, that can be used in the study and design of new promising high-performance Rare-Earth free/lean permanent magnets. The results therein contained might provide some insights into the ongoing debate about the theoretical performance limits beyond Rare-Earth based magnets. Finally, some general strategies are discussed to design possible experimental routes for exploring most promising theoretical novel materials found in the database.  
\end{abstract}

\begin{keyword}
Database \sep magnetic materials \sep permanent magnets \sep Materials Genome Initiative \sep High-throughput \sep VASP \sep Computer simulation \sep NOVAMAG
%\MSC[2010] 00-01\sep  99-00
\end{keyword}

\end{frontmatter}

%\linenumbers

\section{Introduction}

Permanent magnets (PMs) are materials with an internal structure capable of creating an external magnetic field by themselves. Nowadays, these materials play an important role in critical sectors of our advanced society as transport, energy, information and communications technology \cite{Gutfleisch2011}. The magnetic field source given by PMs is widely used in many technological applications, e.g. for inducing mechanical motion in electric motors, making sound in loudspeakers, generating electrical energy in wind turbines, data storage in computer engineering, magnetic resonance imaging in healthcare industry, holding, oil dewaxing, etc \cite{Coeybook,Lewis2013}. In many of these technologies, most efficient designs are frequently achieved using PMs. For instance, the omission of gearboxes in wind turbines based on PM generators considerably reduces maintenance, service costs and overall material utilization \cite{Anja}. Similarly, PM-based motors have many advantages compared to internal combustion engines including high efficiency, compact size, light weight, and high torque \cite{Gutfleisch2011}.

At macroscopic scale, a magnet is described by extrinsic properties that are deduced from the hysteresis loop and determine its performance. These properties are the maximum energy product (BH)$_{max}$ (related to the maximum magnetostatic energy stored and obtained from the second quadrant of the B(H) hysteresis loop), saturation magnetization M$_s$ (total magnetization that is achieved in a strong external magnetic field H that aligns all magnetic domains), remanence M$_r$ (net magnetization after H is removed) and coercivity H$_c$ (magnetic field that reverses a half of the magnetization reducing the overall magnetization to zero,  it determines the "resistance" against demagnetization by external
magnetic fields). The larger (BH)$_{max}$, the smaller amount of material is needed to generate a required magnetic field strength, so high values of (BH)$_{max}$ are desired. Energy product and coercivity depend on the magnet's shape due to its own demagnetizing field. The theoretical upper limit of energy product (BH)$_{max}=\mu_0M_r^2/4$, where $\mu_0$ is vacuum permeability, can be ideally reached for a magnet shape with demagnetizing factor $D=0.5$ and $H_c>M_r/2$ \cite{Skomskibook}. Extrinsic properties also depend on temperature, in fact they typically decrease as temperature increases, reducing the magnet's performance, especially close to the Curie temperature T$_C$ (i.e. ferromagnetic-paramagenetic transition). 

The macroscopic behavior is tightly linked to the microscopic properties called intrinsic. Main magnetic intrinsic properties are atomic magnetic moment $\mu_{at}$ (the magnetic moment per volume gives the maximum theoretical M$_s$), exchange interactions J$_{ij}$ (which determine the magnetic order and T$_C$) and magnetocrystalline anisotropy $K_1$ (that can enhance H$_c$ and it is indispensable in modern magnets to get $H_c>M_r/2$)\cite{Coey2016}. PMs should have high atomic mangetic moments per volume ($>0.1\mu_B/\AA^3$), strong ferromangetic exchange interactions (able to give $T_C>600$ K) and high easy axis magnetocrystalline anisotropy ($K_1>1$ MJ/m$^3$) in order to exhibit good extrinsic properties suitable for PM applcations. In particular, magnetic materials with hardness parameter $\kappa=\sqrt{K_1/(\mu_0M_s^2)}>1$ (called "hard" magnets) are very valuable since can be used to make efficient magnets of any shape \cite{Coey2016}. At mesoscopic scale, intergranular structure between the grains and crystallographic defects can strongly affect the performance of a magnet \cite{Kron}. Therefore, the optimization of the material's microstructure is also very important in the design and development of PMs, especially to approach the theoretical upper limit of (BH)$_{max}$.

Presently, a magnetic material needs to have quite demanding extrinsic properties in order to be considered as a high-performance PM. Desirable values might be $\mu_0M_s>1$ T, $\mu_0H_c>$1 T and (BH)$_{max}>200$ kJ/m$^3$, with cost lower than 100 $\$$/kg \cite{Coey2014}, but it really depends on the specific application \cite{Gutfleisch2011}. Developing new magnets that fulfill these criteria is quite difficult. For example, it took around 1 century to go from modest Co-steel magnets with (BH)$_{max}$=6 kJ/m$^3$ (old fashioned horseshoe shape) to current Rare-Earth (RE) based magnets like sintered Nd$_2$Fe$_{14}$B with (BH)$_{max}$=470 kJ/m$^3$, which is close to its theoretical upper limit 515 kJ/m$^3$. Note also that (BH)$_{max}$ has not improved much in the past 20 years. The great performance of RE-based PMs makes them essential in many technological applications, leading to a strong dependency on expensive RE elements like Nd, Sm or Dy that are monopolized by China's market. This situation has forced PM industry to search for other viable alternatives  based on RE-free PMs as MnAl(C) $\tau$-phase, FeNi L1$_0$, non-cubic FeCo alloys, MnBi, $\alpha''$-Fe$_{16}$N$_2$, exchange spring PM, etc, and also on RE-lean PMs as NdFe$_{12}$ \cite{McCallum,Kuzmin:2014,Skokov:2018}.

Aiming to find new clues, the experimental exploration of new PMs begins to be assisted and guided by computational approaches thanks to their advances in calculation speed, accuracy and reliability \cite{Drebov,Korner}. In 2011, USA launched the Materials Genome Initiative (MGI) alongside the Advanced Manufacturing Partnership to help businesses discover, develop, and deploy new materials twice as fast \cite{MGI_1,MGI_2}. As a result, large open material databases, like AFLOW \cite{Aflow_1,Aflow_2} and Materials Project \cite{Mat_Proj_1,Mat_Proj_2}, have been created, providing a powerful tool for discovering and designing novel materials through machine learning and data mining techniques \cite{LU2017191,LIU2017159}. In Europe, NOMAD repository \cite{nomad_coe} was established to host, organize, and share materials data. At present, NOMAD contains ab-initio electronic structure data from Density Functional Theory (DFT) and methods beyond. Some examples of material properties that have been calculated by high-throughput computational approach are elasticity \cite{Jong_1}, piezoelectricity \cite{Jong_2} and electrolytics \cite{Qu}. A review of some material databases can be found in Ref. \cite{Rev_database}. Concerning to magnetic materials, we highlight the MAGNDATA database \cite{Magndata}, that provides more than 500 published commensurate and incommensurate magnetic structures, and MagneticMaterials.org, an online repository of auto-generated magnetic materials databases that makes use of the ChemDataExtractor toolkit to extract magnetic properties automatically from scientific journal articles and theses, e.g.  recently an auto-generated materials database of T$_C$ has been created using this approach \cite{data_Tc}.

Current databases based on MGI like AFLOW or Materials Project provide information of atomic magnetic moment, but other relevant magnetic properties are missing, like magnetocrystalline anisotropy energy (MAE), exchange integrals, T$_C$, coercivity, energy product, etc which are needed for screening and discovering new high-performance Rare-Earth free/lean PMs. In this work, we describe the Novamag database which aims to complement and extend these databases for the case of PMs by making publicly available many theoretical and experimental results of magnetic materials studied in the H2020 European project Novamag \cite{novamag_web,novamag}. This paper is organized as follows: Section \ref{section:methodology} explains the software technologies used in the development of the database, while Section \ref{section:gui} gives a detailed tutorial on how to use it. Next, Section \ref{section: data} shows a general overview of the available data and the methods used to obtain them. In Section \ref{section:strategy}, we discuss about possible ways to take advantage of Novamag database for exploring the novel promising phases uploaded into it. Finally, this paper ends with a summary of the main conclusions.

\section{Methodology}
\label{section:methodology}

\begin{figure}[h!]
\centering
\includegraphics[width=\columnwidth ,angle=0]{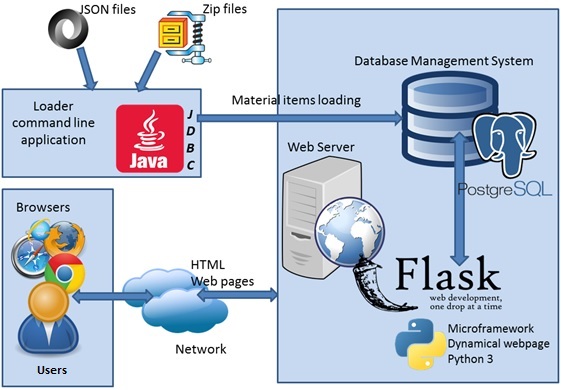}
\caption{Workflow of the Novamag database.}
\label{fig:DB_flow}
\end{figure}

In this section the technologies involved in the development of Novamag database are described. One mandatory requirement present in each decision in the design and implementation of this database is that every component has to be free and widely supported by the community. The application consists of two main modules: i) a web application that uses Phyton ecosystem technologies to show and search the materials and their related information and ii) a data loader application that uses Java ecosystem technologies to insert new items in the database (i.e., NOVAMAG Java Loader Application). These new items are in JSON format. Fig. \ref{fig:DB_flow} shows the general workflow of the Novamag database. Both developments are publicly available in two GitHub projects. The Web Application is in Ref. \cite{web}, while the Java data loader project is in Ref. \cite{loader}. These applications share the same PostgreSQL database \cite{postgresql}. PostgreSQL is the chosen database management system because the following reasons:

\begin{itemize}
\item	The database schema and data items relationships are simple enough to be represented by a relational system. In fact, the schema consists of a main table (i.e., items) containing almost all information about the materials, and a very small set of tables containing information about chemical composition, attached files, and authors.

\item	Relational databases are the most extended database systems, and PostgreSQL is one of the most popular. PostgreSQL is an open source database. It has a graphic easy-to-use administration interface and advanced features that are used in the project (i.e., it supports JSON data, triggers, stored procedures and functions). PostgreSQL presents a high level of conformance to SQL standard ANSI/ISO specifications.

\end{itemize}

The open Novamag database has a Free and Open Source Software (FOSS) licensing and the link to access it is given by Ref. \cite{novamag}.

%------------------------------
%------------------------------ 

\subsection{The Novamag Java Loader Application}

This application is a utility that can read JSON files containing information about one or several materials and load it into the database. The JSON file can reference some attached files, which are supposed to be located in the same path as the JSON file. The application also can cope with zip files containing a subdirectories tree. This subdirectory tree can contain several JSON files and their attached files. Again, the attached files of a JSON file must be located in the same path as the JSON file.

The loader application is invoked by the database administrator from the operative system command line specifying the path of the JSON or ZIP file as a command line argument. Because it is a Java application, it can be invoked both within Unix or Windows operative systems. If remote execution for the database administrator is needed, FTP with SSH for Unix, or a Windows remote terminal for Windows, can be used.
JSON format was adopted because the data is presented in form of aggregation. An item can contain several file references, some arrays like the magnetocrystalline anisotropy energy, or some more complex structures like the atomic positions. XML and JSON are the most extended solutions for this level of complexity in the data representation. XML is a much heavier format than JSON, therefore JSON was considered as the best choice. Presently, efforts are being made to develop a common format for Computational Materials Science Data based on JSON file \cite{nomad_coe,fhi}.

Nowadays, Java is one of the most widespread programming languages. Given that the Loader Application source code is free and available for the community, a very popular language was needed. Application portability for Linux and Windows without recompilation is another issue that was taken into account when Java was chosen. Java is an Object Oriented language. It is very mature and has a lot of libraries and resources. Once Java was chosen for the development of the loader, many of the rest of technologies were determined by such election (i.e., JDBC, JNDI and JUnit).

A logger was used to record the messages the application throws into a log file. Some messages can be eventual error messages, and some others can be confirmations that everything was performed properly. Apache Log4j is one of the most extended loggers for Java, Log4j2 is its newest version. It is a software from Apache Foundation, so its support over the next years is guaranteed. Even being Log4j a very popular logger, SLF4J was used as facade increasing the compatibility with other loggers. That is, the Java loader program only interacts with SLF4J. Then SLF4J receives the Java loader program invocations and translate them into Log4 invocations. In this way, the application does not depend on Log4j and can work in the future with any other logger supported by SLF4J.

Finally, Eclipse was adopted as IDE (Integrated Development Environment) because it is probably the Java most extended IDE. The Mars version was the newest version at the beginning of the development of this application.

\subsection{The Novamag Web Application}

The web application allows to query the information available from the previous importation process. It has been developed using a web framework named Flask.

Flask is a microframework for Python based on Werkzeug and Jinja 2, and it is BSD licensed.  Python is one of the language with greatest growing in the last years, as can be consulted in the Tiobe Ranking (see Ref.  \cite{tiobe}).  Their features make it one of the most powerful programming languages including structured, object-oriented and functional paradigm. On the other hand, the Flask and Python architecture, allows adding new modules in the development in a very simple way.
 
The dynamic web pages are developed using the Model-View-Presenter design pattern using the following Python/Flask libraries:

\begin{itemize}

\item Werkzeug:  collection of various utilities for Web Server Gateway Interface (WSGI) applications and has become one of the most advanced WSGI utility modules (Presenter).  WSGI is a simple calling convention for web servers to forward requests to web applications or frameworks written in the Python programming language.
\item Jinja 2:  a modern and designer-friendly templating language for Python, modelled after Django’s templates. It is fast, widely used and secure with the optional sandboxed template execution environment (View).
\item SQLAlchemy: is the Python SQL toolkit and Object Relational Mapper that gives application developers the full power and flexibility of SQL. It gives data access model with generated SQL from the persistent model in the database (Model).

\end{itemize}

The web requests are processed by a Werkzeug class, that loads the data and redirect the request to the appropriate Jinja2 template, that renders the HTML dynamically using the previous data. The data are loaded using SQLAlchemy code. All the SQL code is automatically generated from a persistent model extracted from the PostgreSQL schema to Python code, using a tool named sqlacodegen (see Ref. \cite{pypi}). All the developed code follows this well-known pattern, allowing an easy maintenance and evolution of the code in the future. 

From the point of view of the web design, it has been loaded the Flask-Bootstrap package. Bootstrap is an open source toolkit for developing with HTML, CSS, and JS providing a standard and responsive web design. This guarantees a correct and uniform visual design in all the web pages.

All this software has been developed using the PyCharm IDE (Integrated Development Environment), version Professional 2017.3, built on November 28, 2017.  The used technologies are all multi-platform, since the product has been developed on Windows, but it is currently deployed and running on Linux. Table \ref{tab:tech_DB} provides a summary of the main technologies used in the Novamag database and their role.

%--------------------------------

\begin{table}[]
  \centering
  \begin{adjustbox}{max width=\textwidth}
\begin{tabular}{|l|l|l|}
\hline
\rowcolor[HTML]{ECF4FF} 
Technology                                                          & Role                                                                                                                                                      & More information                                                          \\ \hline
PosgreSQL v9.2.3                                                    & \begin{tabular}[c]{@{}l@{}}Relational database to store data\end{tabular}                                                                             & \cite{postgresql} \\ \hline
Python 3.0                                                          & \begin{tabular}[c]{@{}l@{}}Programming language used in Flask\end{tabular}                                                                            & \cite{python}                            \\ \hline
Flask 0.12.2                                                        & \begin{tabular}[c]{@{}l@{}}Microframewok to build dynamical webpages\end{tabular}                                                                     & \cite{flask} \\ \hline
\begin{tabular}[c]{@{}l@{}}Flask-Bootstrap 3.3.7.1\end{tabular} & \begin{tabular}[c]{@{}l@{}}Flask update for the design of the webpage\end{tabular}                                                                    & \cite{pythonhosted} \\ \hline
\begin{tabular}[c]{@{}l@{}}Flask-SQLAlchemy 2.3\end{tabular}    & \begin{tabular}[c]{@{}l@{}}Flask update (Object-Relational Mapping)\end{tabular}                                                                      & \cite{flask2} \\ \hline
PgAdmin III                                                         & \begin{tabular}[c]{@{}l@{}}Tool to control PostgresSQL\end{tabular}                                                                                   & \cite{postgresql}                                               \\ \hline
Psql                                                                & Console of PostgreSQL                                                                                                                                     & \cite{postgresql}                                              \\ \hline
PyCharm 2017 2.3                                                    & Integrated development environment for Python                                                                                                             & \cite{jetbrains}                                       \\ \hline
JSON                                                                & Input data format to import data into the database                                                                                                        & \cite{ecma} \\ \hline
Java 8                                                              & \begin{tabular}[c]{@{}l@{}}Programming language used in the application to\\   import JSON data into PostgreSQL database\end{tabular}                     & \cite{oracle}         \\ \hline
JDBC 3                                                              & \begin{tabular}[c]{@{}l@{}}API to program SQL insertions from Java into the\\   database in the data importation application\end{tabular}                 & \cite{oracle2}              \\ \hline
JNDI                                                                & \begin{tabular}[c]{@{}l@{}}Java Naming and Directory Interface to store database \\    connection settings in the data importation application\end{tabular} & \cite{oracle3}    \\ \hline
Apache Log4j2                                                       & Logger in the data importation Java application                                                                                                           & \cite{apache}                                    \\ \hline
SLF4J                                                               & \begin{tabular}[c]{@{}l@{}}Simple Logging Façade for the data importation Java\\   application\end{tabular}                                               & \cite{slf4j}                                                   \\ \hline
JUnit 4                                                             & \begin{tabular}[c]{@{}l@{}}Testing framework for Java used in the data\\   importation application\end{tabular}                                           & \cite{junit} \\ \hline
GitHub                                                              & \begin{tabular}[c]{@{}l@{}}Version control repository used both for the Java\\   data importation application and for the web application\end{tabular}    & \cite{github} \\ \hline
Eclipse Java Mars                                                   & \begin{tabular}[c]{@{}l@{}}Integrated development environment used in the\\   project for Java programming\end{tabular}                                   & \cite{eclipse} \\ \hline
\end{tabular}
\end{adjustbox}
  \caption{Summary of the main technologies used in Novamag database.}
  \label{tab:tech_DB}
\end{table}

%-------------------------------

\subsection{Uploading procedure}

\begin{figure}[h!]
\centering
\includegraphics[width=\columnwidth ,angle=0]{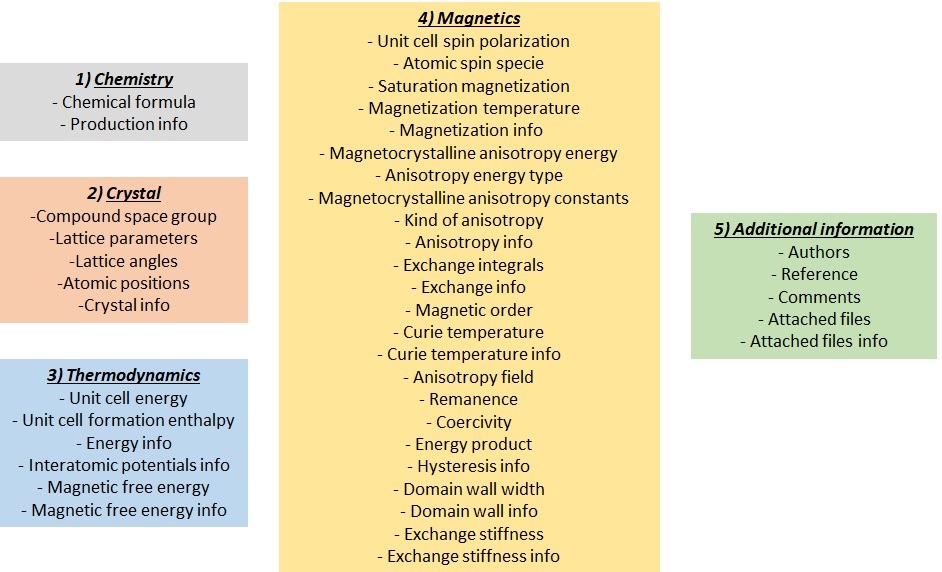}
\caption{List of theoretical and experimental properties included in Novamag database.}
\label{fig:properties}
\end{figure}

In the first stage of the database development, a set of theoretical and experimental properties to be included in the database was agreed by the Novamag consortium. Fig. \ref{fig:properties} shows the full list of properties. These properties are classified into 5 categories following MGI standards and the Review of Material Modelling (RoMM) vocabulary: (i) chemistry, (ii) crystal, (iii) thermodynamics, (iv) magnetics and (v) additional information. 
%------------------------------
%------------------------------ 

In order to deal and upload the large amount of data generated by computational high-throughput approaches (see Sections \ref{section: AGA} and \ref{section:mae}), we created an automated transfer procedure using metadata JSON files written by bash scripting. Once the data is in JSON format it can be uploaded into the database by the NOVAMAG Java Loader Application. A JSON file can contain information of a solely material or of several materials. The attached files of these materials must be allocated in the same path than the JSON file for the loading. The NOVAMAG Java Loader needs several jar libraries that provide methods to manage JSON format, JDBC database connectivity, JNDI services and logging. If a large data upload is needed, then Java Loader Application can cope with ZIP compressed files. The ZIP files can contain a sub-directory tree. In each sub-directory new sub-directories or JSON files can be allocated.

\section{Graphical User Interface}
\label{section:gui}

\begin{figure}[h!]
\centering
\includegraphics[width=\columnwidth ,angle=0]{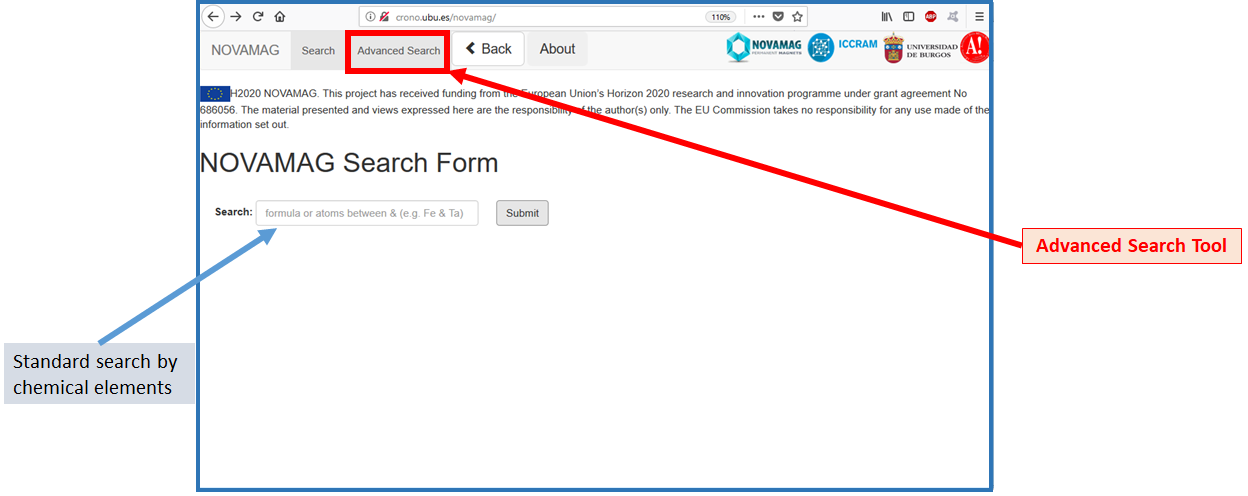}
\caption{Screenshot of the home webpage of Novamag database GUI.}
\label{fig:GUI_1}
\end{figure}

A friendly Graphical User Interface (GUI) has been developed to use and explore the Novamag database easily. Fig. \ref{fig:GUI_1} shows the home webpage of Novamag database, which contains two search modes: i) standard and ii) advanced.

%------------------------------
%------------------------------ 

The standard mode allows to perform quick and general search through the database by specifying the chemical name abbreviation\cite{novamag}. For example, if one wants to search for compounds with iron and nickel, then one should type "Fe \& Ni" inside the box called "Search" and then click on the button "Submit", see Fig. \ref{fig:GUI_1}. In this case, the system will show all available compounds that contain the specified chemical elements. Additionally, one can search for specific structures by typing the chemical formula, e.g. "FeNi2".

\begin{figure}[h!]
\centering
\includegraphics[width=\columnwidth ,angle=0]{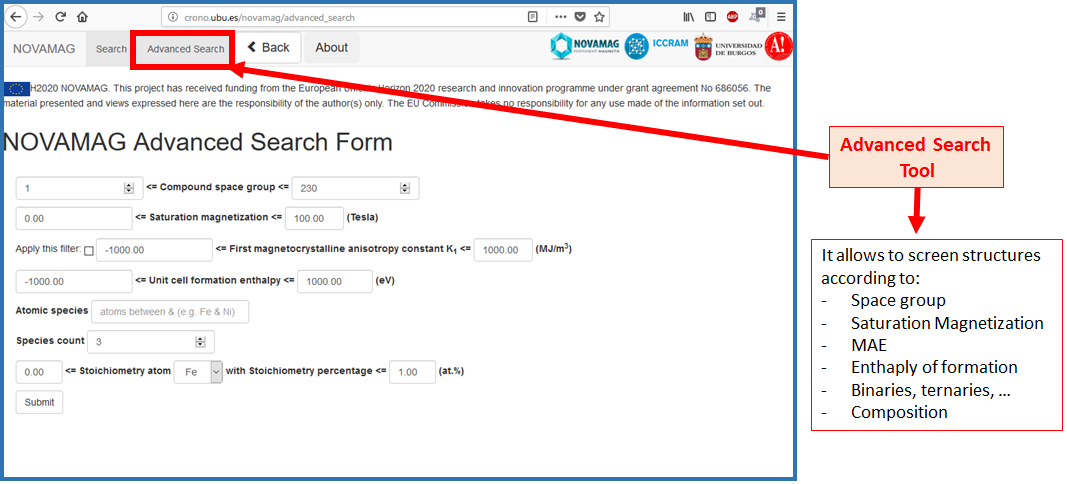}
\caption{Screenshot of the advanced search tool of Novamag database GUI.}
\label{fig:GUI_2}
\end{figure}

Alternatively, users can also perform advanced search by clicking on the button “Advanced Search” \cite{novamag2}, that is placed on the top of home webpage, see Fig. \ref{fig:GUI_1}. Next, after doing that, it will appear the advanced search site, see Fig. \ref{fig:GUI_2}. At present, it allows to filter and screen compounds according to: Crystallographic space group symmetry, saturation magnetization (M$_s$), first magnetocrystalline anisotropy constant (K$_1$), unit cell formation enthalpy, atomic species, species count and stoichiometry.

%------------------------------
%------------------------------ 

These filters are very useful to identify promising materials for permanent magnet development in a first screening stage. To this end, we recommend to select uniaxial space groups (from 75 to 194) because they might exhibit a well-defined magnetocrystalline anisotropy easy axis, saturation magnetization above 1 Tesla, negative unit cell formation enthalpy (in order to reject very unstable phases), atomic species like magnetic elements (Fe, Ni, Co) and non-Critical Raw Materials (that could be light elements to induce tetragonal distortions and phase stabilization like B, C, N, P, etc or/and transition metals with high spin-orbit coupling as Ta, Hf, etc), species count (binaries, ternaries, etc) and stoichiometry corresponding to Fe-rich compounds (with Fe content above 50 \% at.) to enhance saturation magnetization and Curie temperature. Additional filters as Curie temperature, coercivity or maximum energy product can also be useful in a second screening stage, and might be implemented in the future. Some general search strategies through the database are discussed in Section \ref{section:strategy}.

%------------------------------
\begin{figure}[h!]
\centering
\includegraphics[width=\columnwidth ,angle=0]{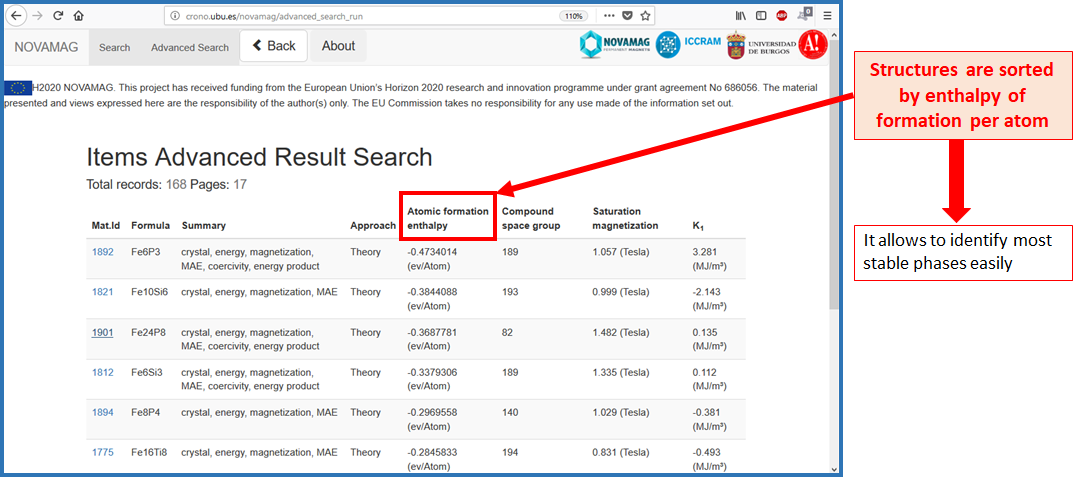}
\caption{Screenshot of Novamag database GUI showing the list of materials found in an advanced search. Items are sorted by enthalpy of formation per atom, so that most stable phases can be identified easily.}
\label{fig:GUI_3}
\end{figure}
%------------------------------  

Once the search settings are ready and submitted, then the list of items that fulfill these criteria are presented, see Fig. \ref{fig:GUI_3}. By default, items are sorted by enthalpy of formation per atom, so that most stable phases for a given chemical formula can be identified easily. In the list of items, it is shown some properties or fields of each compound in different columns like: material identification number (Mat.Id), chemical formula, summary (data available of each item), approach (theory or experiment), atomic formation enthalpy, space group, saturation magnetization and first magnetocrystalline anisotropy constant.

\begin{figure}[h!]
\centering
\includegraphics[width=\columnwidth ,angle=0]{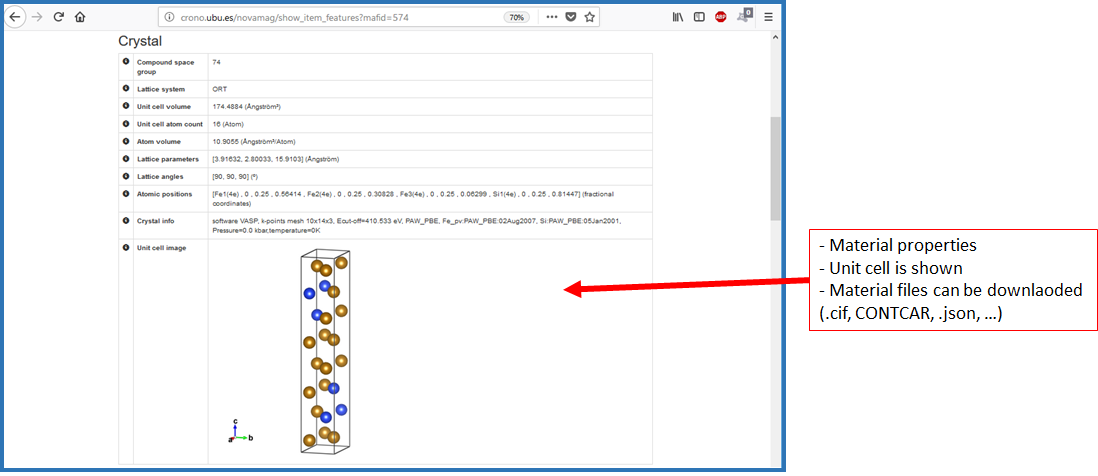}
\caption{Screenshot of Novamag database GUI showing material properties of a selected item.}
\label{fig:GUI_4}
\end{figure}

In order to see more details of a compound that is in the list one should click on the corresponding material identification number (called Mat.Id, at the first column). Then, all available data of this compound will be shown, see Fig. \ref{fig:GUI_4}. On the left hand side of the name of each field or property, there is a small black icon with label "i", by clicking on it one can see more information about the definition and physical units of that property. Finally, material metadata files like crystallographic information file (.CIF), JSON file used in the database to upload it, CONTCAR (output file of VASP code that describes the relaxed unit cell), figure of the unit in portable network graphics format (.png), etc can be found and downloaded in section "Additional information" (at the end of the webpage of each material).
 
%------------------------------
%------------------------------ 

\section{Material Data Available}
\label{section: data}

 This section gives a brief overview of the data available in Novamag database. Additional theoretical and experimental contributions from the Novamag consortium, not shown or mentioned here, are expected to be uploaded into the database in the near future.

\subsection{Theoretical structures calculated with an Adaptive Genetic Algorithm}
\label{section: AGA}

One of the most important and difficult step in the computational high-throughput  technique for materials design is the prediction of new stable crystal phases. Here, we make use of an  Adaptive Genetic Algorithm (AGA) \cite{Oganov} to explore the crystal phase space for many Rare-Earth-free/lean magnetic binary and ternary compounds. This task was computationally performed using the software USPEX \cite{uspex} combined with Vienna Ab Initio Simulation Package (VASP) \cite{vasp_1,vasp_2,vasp_3}. USPEX uses as an input the number and type of ions to be considered within the unit cell only. At the first step, a set of structures is generated at random, by randomly choosing a crystal space group, corresponding lattice vectors, and ion positions. These initial structures are far from their equilibrium, thus performing a structure relaxation is required to estimate accurately the energy of each one, which serves as a fitness criterion. A subset of fitted structures is selected to generate a next generation of structures by means of genetic operations (crossover and mutations). The process of random generations of structures is performed also at each step to provide a diversity of structures for each generation. The search for an optimal structure is performed until no new best structures are generated for a certain number of generations or the maximum number of generations is reached. We modified and optimized the interface between USPEX and VASP codes in order to improve the performance of structural optimization as well as to perform calculations in a high-throughput manner. \cite{Arapan_AGA_1,Arapan_AGA_2,Arapan_AGA_3}
 
%------------------------------
\begin{figure}[h!]
\centering
\includegraphics[width=\columnwidth ,angle=0]{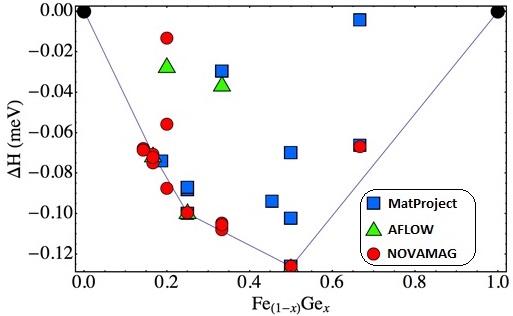}
\caption{Calculated formation enthalpy at zero-temperature for some low-energy Fe-Ge metastable phases obtained in Novamag using AGA (red circle), and found in Material databases: Materials Project (blue square) and AFLOW (green triangle). Blue line represents the enthalpy hull line.}
\label{fig:FeGe}
\end{figure}
%------------------------------  

In Novamag project, over 15,000 structures have been calculated with AGA methodology, where we mainly explored Fe-rich binary and ternary compounds without RE-elements. Typically, more than 100 of structures are calculated for a given chemical formula. At the moment,  only around 10-20 most stable phases (i.e. with the lowest energy) per chemical formula are being uploaded to the Novamag database. The formation enthalpy of these structures is calculated as (binary phase)
%%%%%%%%%%%%%%%%%%%%%%%%%
\begin{equation}
\Delta H_{F}(X_aY_b)=E(X_aY_b)-a\cdot E(X)-b\cdot E(Y),
\end{equation}
%%%%%%%%%%%%%%%%%%%%%%%%
where $E(.)$ is the energy, $a$ and $b$ are the number of atoms of $X$ and $Y$ in the formula unit $X_aY_b$, respectively. Fig. \ref{fig:FeGe} presents calculated formation enthalpy at zero-temperature for some Fe-Ge metastable phases obtained in Novamag using AGA, and found in MGI databases like Materials Project\cite{Mat_Proj_1,Mat_Proj_2} and AFLOW\citep{Aflow_1,Aflow_2}. We see that AGA is able to find many structures close the enthalpy hull line (blue line), reproducing and extending the data available in other databases based on MGI.

The amount of generated data of each structure depends on the number of atoms in unit cell and on the number of AGA generations. We may consider as representative a compound with 12 atoms/unit cell, which may generate up to 100GB of data for 650 structures, or about 150MB/structure. We can estimate the overall amount of generated data to be about 2TB, that are stored and backed-up in office computers, external hard drives and Relational Database (PosgreSQL). A back-up of all data is done every month.

\subsection{High-throughput search for structures with high magnetocrystalline anisotropy}
\label{section:mae}

One of the key intrinsic properties of modern PMs is MAE because it can greatly increase the coercivity. In this section, we discuss about three computational high-throughput strategies performed in the Novamag project for finding magnetic structures with high easy-axis MAE. 

\subsubsection{Screening of metastable uniaxial phases from AGA and databases}
\label{section:mae_AGA}

Only a small amount of phases from the large set of theoretical uniaxial magnetic structures close the enthalpy hull line might exhibit a high easy-axis MAE. Hence, a computational high-throughput approach is needed in order to analyze them and be able to identify most promising ones. To this end, we devised a script that controls the calculation of MAE in an automated way providing as the input only the atomic position file. For each structure, calculations are performed in several steps: full structural optimization (volume, cell and atomic positions) with VASP standard precision, followed by more accurate calculations with optimal code parameters obtained for a set of representative known permanent magnets. For the optimized structure, accurate charge density and wave function are calculated for ferromagnetic arrangement of spins via collinear spin-polarized DFT calculations, which will serve as input for non-collinear spin calculations (NCL) with spin-orbit coupling. NCL-calculations are performed in a non-self-consistent manner for a set of polar and azimuthal angles. This procedure is fed by the crystallographic data of Rare-Earth free/lean theoretical structures predicted by AGA (section \ref{section: AGA}) and found in open material databases that are previously screened according to space group (only uniaxial phases are considered), formation enthalpy ($<$0) and saturation magnetization ($\gtrsim$1T), see Fig. \ref{fig:mae_flow}.

%------------------------------
\begin{figure}[h!]
\centering
\includegraphics[width=\columnwidth ,angle=0]{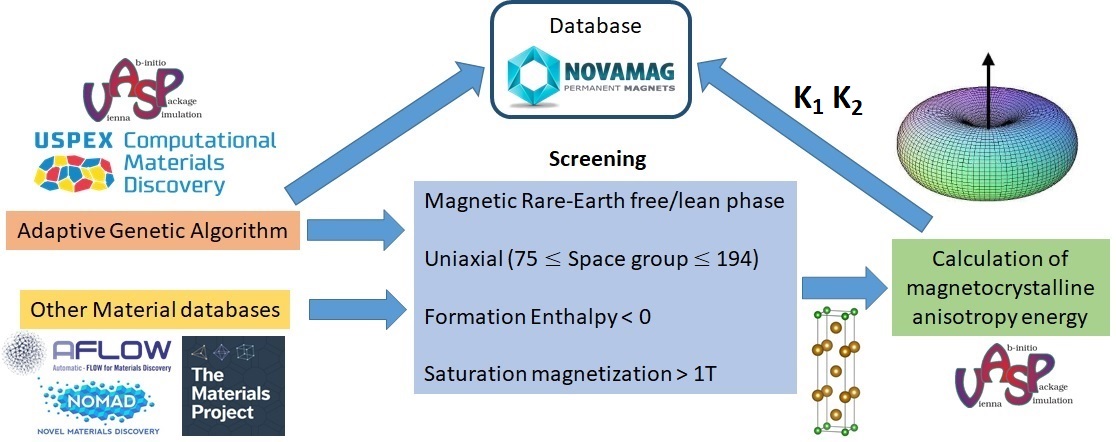}
\caption{Workflow of the high-throughput calculation of magnetocrystalline anisotropy energy developed in Novamag project for the discovery of high-performance Rare-Earth-free/lean permanent magnets.}
\label{fig:mae_flow}
\end{figure}
%------------------------------ 

Once the energy of an uniaxial unit cell is calculated, constraining the atomic magnetic moments at different polar angles $\theta$, then first and second anisotropy constants K$_1$ and K$_2$ are obtained by fitting the energy data to 2nd order function
%%%%%%%%%%%%%%%%%%%%%%%%%%%%%%%%%%%%%%%%%
\begin{equation}
E(\theta)=K_1 \sin^2 \theta+K_2 \sin^4 \theta.
\end{equation}
%%%%%%%%%%%%%%%%%%%%%%%%%%%%%%%%%%%%%%%%  
A final procedure, written by bash scripting, was also prepared to upload the large amount of generated data from this approach to the database automatically. As an example, in Fig. \ref{fig:mae_flow} we show the high-throughput calculation of first anisotropy constant K$_1$ at T=0K as a function of polarization $\mu_0$M$_s$ for some theoretical uniaxial Fe-based binaries (Fe-X where X= Al, B, P and Hf) with $\Delta H_F<0$ and easy axis, predicted by AGA in Novamag project and found in AFLOW database. We observe this approach is capable to reveal hidden RE-free theoretical phases with hardness parameter $\kappa=\sqrt{K_1/(\mu_0M_s^2)}>1$, such as hexagonal P-6m2 FeAl ($\kappa=2.6$), and tetragonal I4/mmm structures Fe$_{16}$B$_{2}$ ($\kappa=1.1$) and Fe$_4$Hf$_2$ ($\kappa=3.0$). Many other Fe-based and Co-based binaries have been calculated and uploaded following this methodology.

%------------------------------
\begin{figure}[h!]
\centering
\includegraphics[width=\columnwidth ,angle=0]{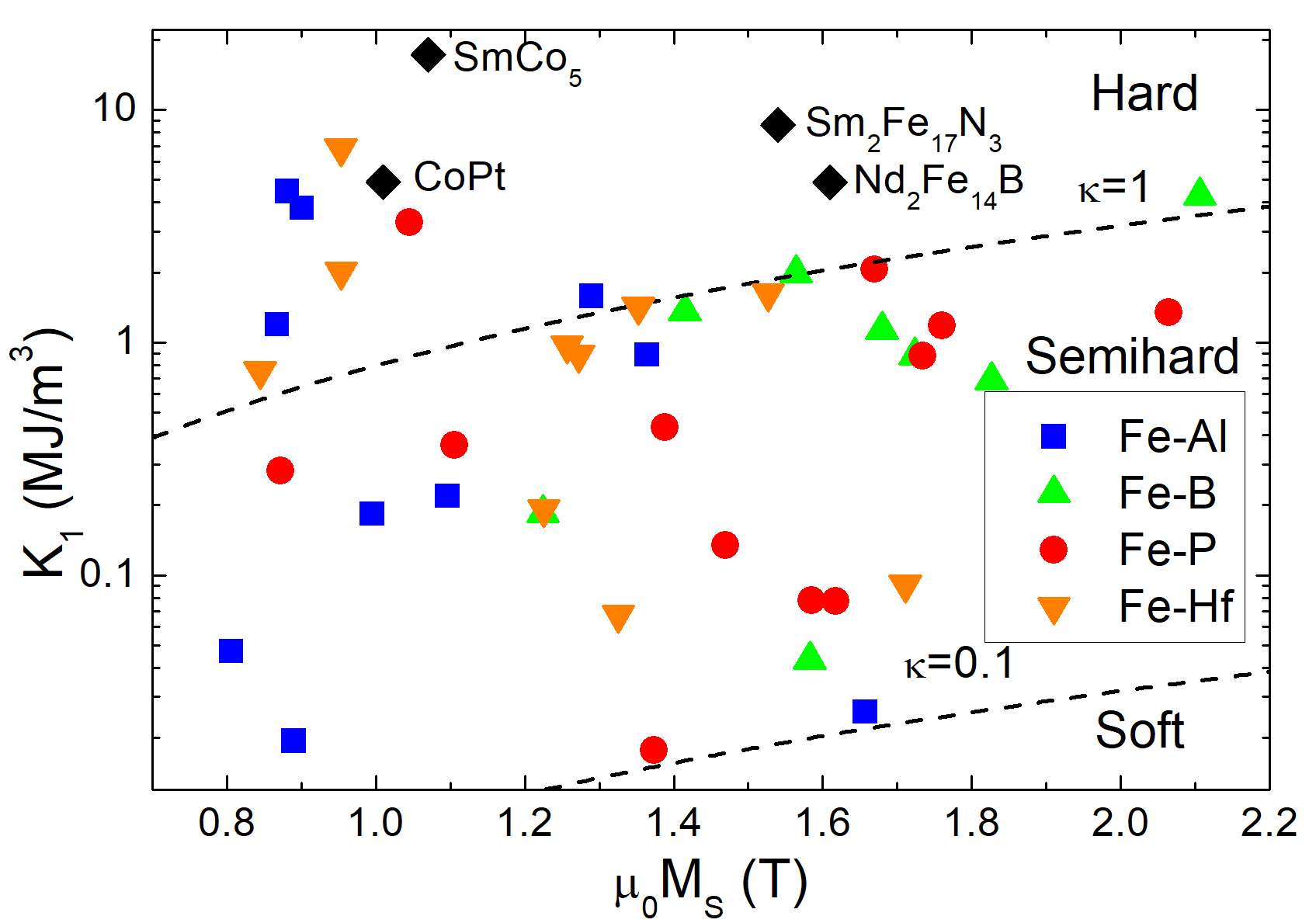}
\caption{High-throughput calculation of first anisotropy constant K$_1$ as a function of polarization $\mu_0M_s$ for some theoretical uniaxial Fe-based binaries (Fe-X where X=Al, B, P and Hf) with $\Delta H_F<0$. The dashed lines correspond to magnetic hardness parameter $\kappa=\sqrt{K_1/(\mu_0M_s^2)}$ for values $\kappa$ =1 and 0.1. Hard magnetic materials ($\kappa >1$) can be used to make efficient magnets of any shape. Black symbols stand for some known hard phases \cite{Coeybook}.}
\label{fig:mae_flow}
\end{figure}
%------------------------------ 

\subsubsection{Tetragonally distortion of cubic phases}
\label{subsection:tetragonal distortion}

\begin{figure}[h]
\centering
\includegraphics[width=\columnwidth ,angle=0]{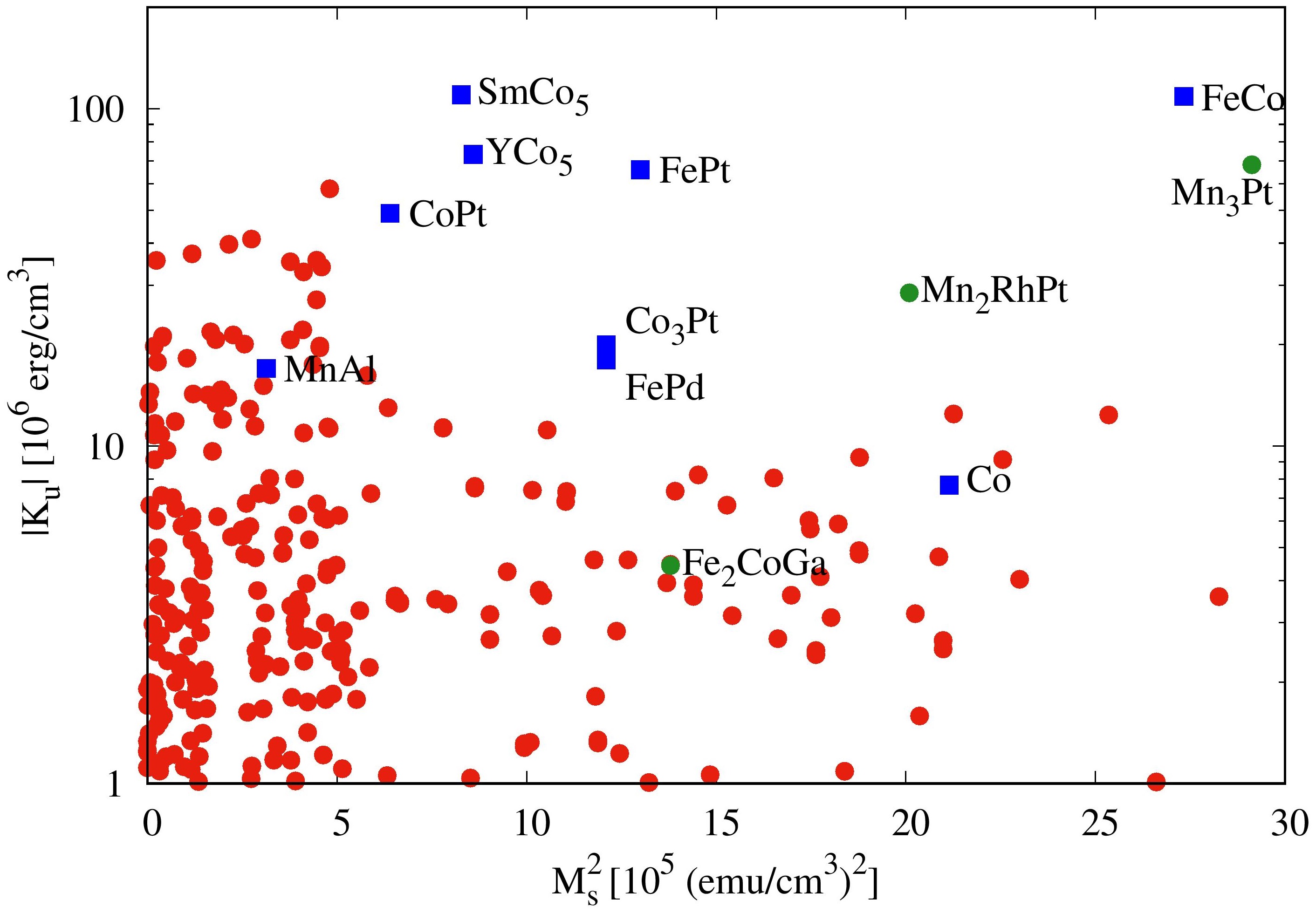}
\caption{High-throughput calculation of the first anisotropy constant K$_1$ as a function of magnetization for stable binary and ternary magnetic materials. 
The blue squares mark the known experimental compounds, and the green circles denote three promising candidates (Mn$_3$Pt, Mn$_2$RhPt, and Fe$_2$CoGa) 
identified as permanent magnets.}
\label{fig:mae_distcubic}
\end{figure}

Enlightened by the work of Burkert {\it et. al.} on tetragonal FeCo alloys,~\cite{Burkert:2004} we have performed high throughput
calculations on the tetragonally distorted cubic magnetic materials to screen for candidates as permanent magnets, as shown in
Fig.~\ref{fig:mae_distcubic}. All the binary and ternary compounds with cubic structures including one of Cr, Mn, Fe, Co, and Ni 
from the Materials Project database~\cite{Mat_Proj_1,Mat_Proj_2} are considered, where 1\% compressive strain is applied to get the tetragonally distorted structures
under the constant volume approximation. There are quite a few compounds showing large magnetocrystalline anisotropy
induced by the imposed tetragonal distortions. For instance, Mn$_3$Pt shows an anisotropy as large as $\vert K_1\vert=$6.8 MJ/m$^3$, and Mn$_2$RhPt with an anisotropy of $\vert K_1\vert=$2.8 MJ/m$^3$. Concerning the sign of $K_1$, it is noted that compressive and tensile strain will lead to opposite magnetocrystalline anisotropies
as the original cubic materials are isotropic to the first order due to the high symmetry. An interesting question is how such tetragonal
distortions can be stabilized. We suspect that light interstitial atoms such as H, B, C, and N will lead to substantial tetragonal distortions,
as recently investigated in Fe~\cite{Zhang:2016} and FeCo~\cite{Reichel:2017, Salikhov:2017} alloys.
Systematic calculations have been performed on compounds with the Cu$_3$Au (as for Mn$_3$Pt) and Heusler (as for Mn$_2$RhPt and Fe$_2$CoGa) structures with light interstitials, and the results will be reported elsewhere.

%------------------------------
%------------------------------ 

\subsubsection{Modified magnetic materials from first principles}
\label{subsection:UPP}

\begin{figure}[h]
\centering
\includegraphics[width=\columnwidth ,angle=0]{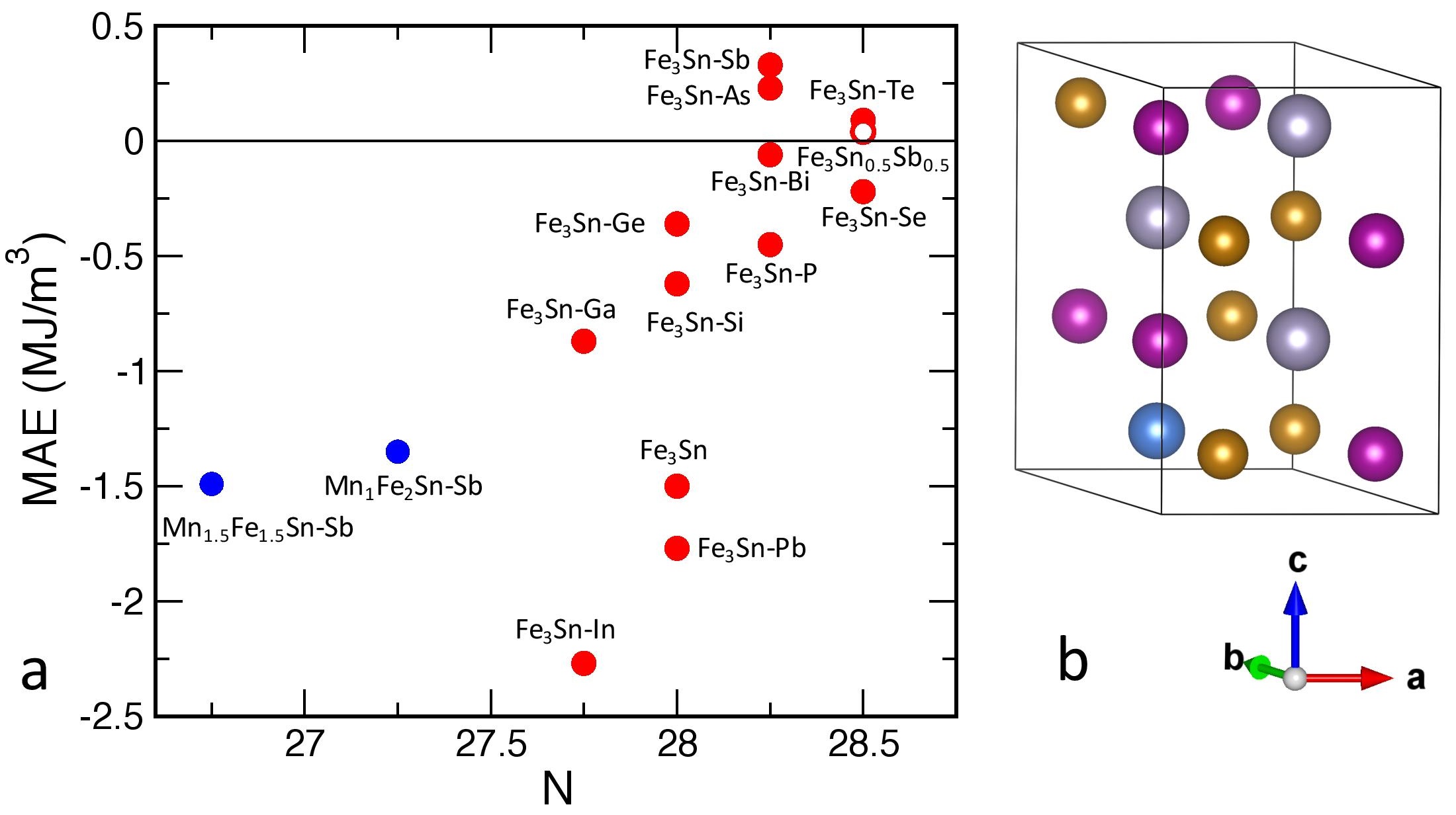}
\caption{MAE and crystal structure of the Fe$_y$Mn$_{3-y}$Sn$_x$M$_{1-x}$ system, y=1.5, 2, and 3; x= 0.5, 0.75, and 1. a) The magnetocrystalline anisotropy energy (MAE) calculated from first principles as a function of the number of valence electrons, N, per formula unit. 
b) The $1\times1\times2$ Fe$_3$Sn hexagonal cell with one impurity atom on the tin sublattice. Fe atoms are shown with brown spheres, Sn atoms with grey and M (M = Si, P, Ga, Ge, As, Se, In, Sb, Te, Pb and Bi) impurity atom is shown with the blue sphere. Violet spheres represent the Mn atoms on the Fe sublattice.}
\label{fig:mae_fe3sn}
\end{figure}

Another interesting computational high-throughput strategy to find promising PMs is to try different dopants on known magnetic structures in order to tune and improve their properties. For instance, the Novamag database provides data of first principles calculations for the RE-free modified Fe$_3$Sn compound \cite{Olga}. Among all the Fe-based intermetallic compounds the Fe$_3$Sn phase is the most attractive, due to the highest concentration of Fe and therefore high magnetic moment. However, as it has been shown experimentally, the magnetocrystalline anisotropy of this compound is planar \cite{Sales}, while the uniaxial anisotropy is a requirement for a good PM. We have studied the influence of dopants on the magnetocrystalline anisotropy of Fe$_3$Sn (see Fig. \ref{fig:mae_fe3sn} a).  On the Sn sublattice we considered 25 at.\% of impurities, such as M = Si, P, Ga, Ge, As, Se, In, Sb, Te, Pb, and Bi. On the Fe sublattice Mn was added as a structure stabilizer. 
%------------------------------
%------------------------------ 

The high temperature phase of Fe$_3$Sn has a hexagonal crystal structure with space group P6$_3$/mmc ($\#$194). The $1\times1\times2$ supercell of Fe$_3$Sn comprising of 12 Fe and 4 Sn atoms, one of which was substituted by a dopant (see Fig. \ref{fig:mae_fe3sn} b) was considered. In the case of doping with Mn, 6 or 4 Fe atoms were substituted. Structures were relaxed using  VASP \cite{vasp_1,vasp_2,vasp_3} within the projector augmented wave (PAW) method \cite{PAW}. The electronic exchange and correlation effects were treated by the generalized gradient approximation (GGA) in the Perdew, Burke, and Ernzerhof (PBE) form \cite{PBE}. For the calculation of the magnetic properties the full-potential linear muffin-tin orbital (FP-LMTO) method implemented in the Relativistic Spin Polarized toolkit (RSPt code) \cite{RSPT_1,RSPt_2} was used. We performed integration over the Brillouin zone, using the tetrahedron method with Bl\"{o}chl's correction \cite{bloechl_tetra}. The k-point convergence of the MAE for the chosen supercell size was found when increasing the Monkhorst-Pack mesh \cite{monkhorst-pack} to $24\times24\times24$ used in all calculations.

The calculated MAEs were plotted as a function of the number of valence electrons per formula unit for doping atoms on the Sn and Fe sublattices, see Fig. \ref{fig:mae_fe3sn}. Positive numbers correspond to the uniaxial MAE. The addition of certain elements, such as Sb, As, and Te allowed one to change MAE from planar to uniaxial, however the absolute value was rather small. The maximal uniaxial MAE found (for As and Sb dopants), is obtained for Group VA elements of the periodic table. 

Calculated MAE of system with increased Sb content, i.e. Fe$_3$Sn$_{0.5}$Sb$_{0.5}$ with 28.5 valence electrons per formula unit, was found to be close to the value for Fe$_3$Sn$_{0.75}$Te$_{0.25}$ with the same number of valence electrons (see opaque circle in Fig.\ref{fig:mae_fe3sn}). Therefore, for this system the MAE depends more on the number of valence electrons rather than on the dopant. It was shown experimentally \cite{Olga} that  the Fe$_3$Sn$_{0.5}$Sb$_{0.5}$ system is unstable unless doped with Mn. However, adding Mn to Fe$_3$Sn$_{0.75}$Sb$_{0.25}$ system changes anisotropy back to planar. Though the increase of Mn content leads to the increase of absolute value of MAE.

\subsection{Exchange integrals and Curie temperature}

Another important intrinsic property of magnetic materials is the exchange energy, which is typically described by the classical Heisenberg Hamiltonian,
%%%%%%%%%%%%%%%%%%%%%%%%%%%%%%%%%%%5
\begin{equation}
E_{ex}=-\frac{1}{2} \sum_{i,j}J_{ij} \textbf{s}_i\cdot \textbf{s}_j,
\label{eq:Heis}
\end{equation}
%%%%%%%%%%%%%%%%%%%%%%%%%%%%%%%%%%%%%%
where $J_{ij}$ are the exchange parameters and $\textbf{s}_i$ is the unit vector along the magnetic moment of atom i-th. The factor $1/2$ takes care of the double counting in the summation. This is the convention adopted in the database, but note there are many works and codes where Eq. \ref{eq:Heis} is defined without the factor $1/2$, so in these cases the $J_{ij}$ parameters are 2 times smaller than those from Eq. \ref{eq:Heis}. A warning message is displayed in the database's field "exchange info" when a different convention is used in order to avoid any  misunderstanding. From $J_{ij}$ one can determine the magnetic order (ferromagnet, antiferromagnet, etc), roughly estimate the T$_C$ using mean-field approximation (MFA) or random phase approximation (RPA), and build atomistic spin dynamics (ASD) models \cite{Eriksson_book,Kurz,Skubic,uppasd,Evans,vampire} to calculate the temperature dependence of magnetic properties (magnetization, domain wall width, exchange stiffness, etc).

\subsubsection{Rare-Earth-free compounds}

\begin{figure}[h!]
\centering
\includegraphics[width=\columnwidth ,angle=0]{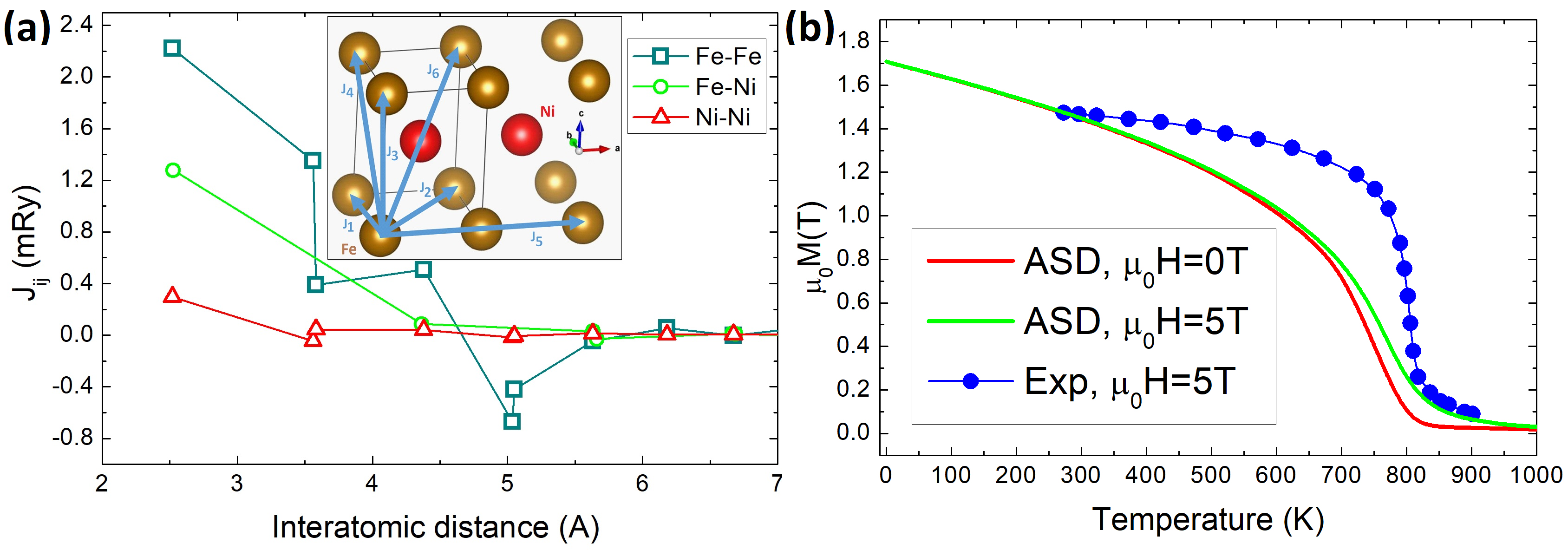}
\caption{(a) Estimated exchange integrals of L1$_0$ FeNi as a function of the interatomic distance. Inset shows a diagram of the Fe-Fe exchange parameters up to the sixth nearest neighbour (labelled by J$_i$ i=1,...,6). (b) Total magnetization versus temperature of L1$_0$ FeNi  with no external magnetic field (red line) and external magnetic field $\mu_0$H=5T (green line), obtained by ASD simulations. Blue circles correspond to experimental data of magnetization with external magnetic field $\mu_0$H=5T reported in Ref. \cite{Lewis}.}
\label{fig:J_FeNi}
\end{figure}

First principles calculations of J$_{ij}$ have been calculated and uploaded to the database for some promising Rare-Earth free PMs such as MnAl $\tau$-phase \cite{Nieves} or L1$_0$ FeNi. For instance, Fig. \ref{fig:J_FeNi}a shows the estimated exchange integrals of L1$_0$ FeNi as a function of the interatomic distance using frozen magnon calculations implemented in FLEUR code \cite{Kurz,Jacob,Lezaic}. Note FLEUR code follows the definition given by Eq. \ref{eq:Heis} that includes factor $1/2$. We employed the spin spiral formalism with 1960 k-points in the irreducible Brillouin zone (IBZ), plane-wave cut-off of k$_{max}$= 4.1 a.u.$^{-1}$ and 463 q-points in the IBZ.  We see that Fe-Fe intralayer exchange interaction is stronger than interlayer one. We also observe that Fe-Ni and Ni-Ni exchange interactions are quite weak, in fact only first nearest neighbour interaction might play a relevant role in the spin dynamics. The Curie temperature given by ASD simulations using these exchange parameters is T$_C$=800 K, which is in good agreement with experiment \cite{Lewis} (see Fig. \ref{fig:J_FeNi}b). The details of this ASD model and calculation's settings can be found in the database. Note a better agreement of magnetization in the all temperature range can be achieved by rescaling the temperature in the classical ASD simulations \cite{Evans2} or including quantum effects \cite{Woo,Kormann}. Additionally, the database contains data about exchange stiffness and domain wall width given by ASD models, which can be useful for micromagnetics simulations (see Section \ref{subsection:micromag}) within a multiscale modelling approach \cite{Kazan}.

%------------------------------
%------------------------------ 

\subsubsection{Rare-Earth-lean 1:12 compounds}

\begin{figure}[h]
\centering
\includegraphics[width=\columnwidth, angle=0]{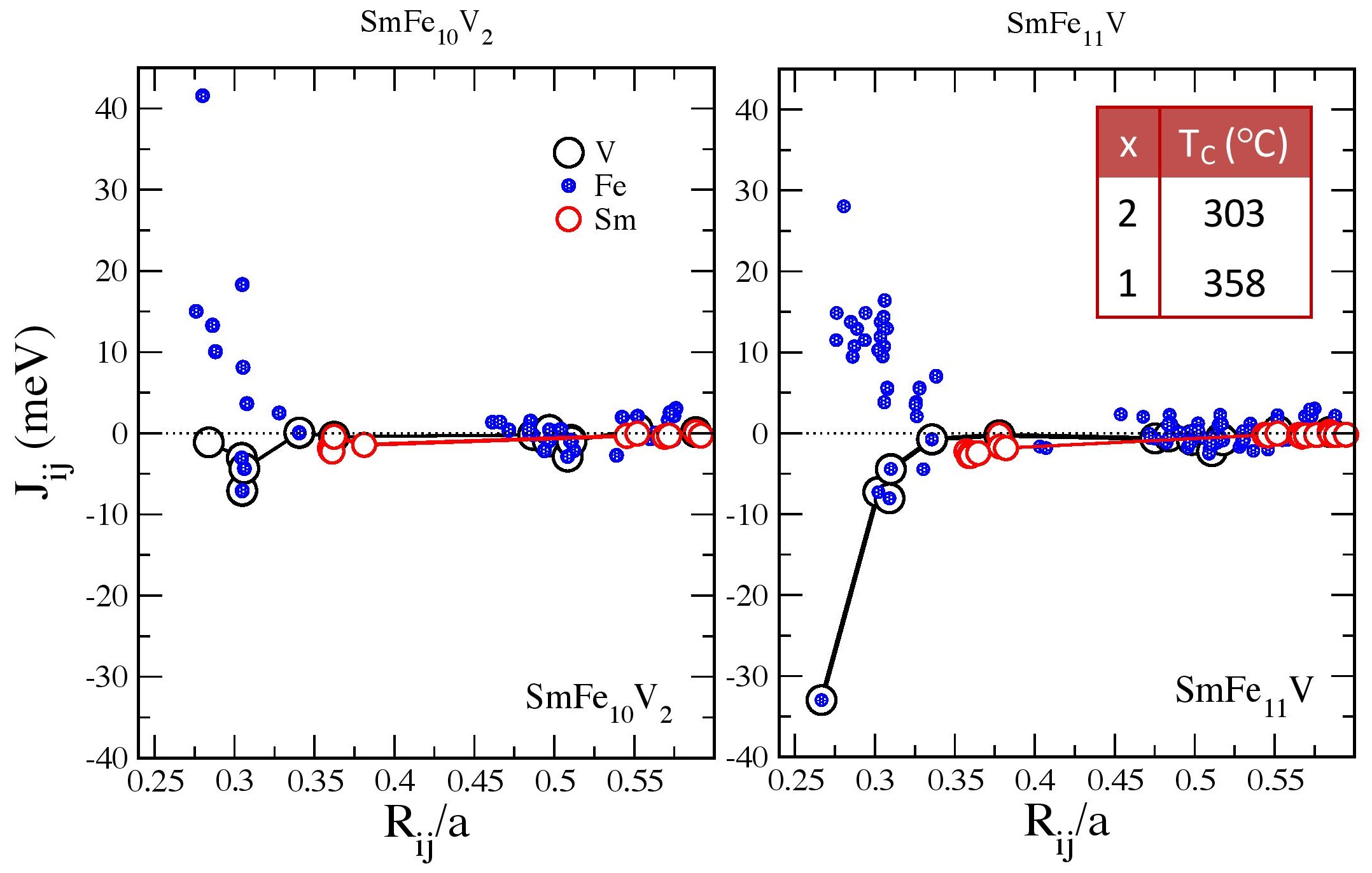}
\caption{Inter-site exchange parameters J$_{ij}$ between atoms $i$ and $j$ separated by the distance R$_{ij}$ of SmFe$_{10}$V$_2$ (a) and SmFe$_{11}$V (b) compounds. The J$_{ij}$s are shown for all R$_{ij}$s involving i=V, Fe, and Sm (shown by large black, small filled blue, and medium red circles, respectively). $a$ stands for the lattice constant. In the inset the mean-field estimation of T$_C$ is given; x stands for the V content, i.e. 2 in  SmFe$_{10}$V$_2$ and 1 in SmFe$_{11}$V.}
\label{fig:J_SmFeV}
\end{figure}

Exchange parameters for the known SmFe$_{10}$V$_2$ system, as well as for the newly stabilized SmFe$_{11}$V \cite{Ana_Maria} were obtained using Lichtenstein's method \cite{Jij_1, Jij_2}, as implemented in FP-LMTO code RSPt \cite{Jij_3}. Note code RSPt defines Eq. \ref{eq:Heis} without factor $1/2$, so the following results are based on the convention used in this code. The electronic exchange and correlation effects were treated by the generalized gradient approximation (GGA) in the Perdew, Burke, and Ernzerhof (PBE) form \cite{PBE}.  The k-point mesh equal to $18\times18\times24$ was used in all calculations. In Fig. \ref{fig:J_SmFeV} the calculated J$_{ij}$s are shown for pairs of atoms $ij$ with $i$ fixed to V, Fe, and Sm, and all corresponding $j$s at distances R$_{ij}$ from the first coordination shell up to 0.6 of the lattice constant. As one may see from Fig. \ref{fig:J_SmFeV}, the strongest exchange interactions are between Fe atoms that fall off with the distance and are close to zero starting at a distance of approximately 0.35 of the lattice constant. They are positive, i.e. favor the ferromagnetic alignment of magnetic moments on the Fe atoms, and therefore play a dominating role in the paramagnetic-to-ferromagnetic transition and define to a large degree the transition temperature (T$_C$). The behavior of the Fe-Fe interactions is qualitatively rather similar in both SmFe$_{10}$V$_2$ and SmFe$_{11}$V compounds, although the highest values differ by up to 50 \% being larger for SmFe$_{11}$V. It is also noticeable that all the V-V as well as all the types of interactions involving Sm are rather weak. The Fe-V interaction at the first coordination shells is, however, relatively large, though around factor 2 weaker than the strongest Fe-Fe interactions in SmFe$_{10}$V$_2$. In both cases they favor the antiferromagnetic alignment of magnetic moments on the Fe and V atoms. In SmFe$_{10}$V$_2$, where there are the Fe-V pairs in the first coordination shell, the antiferromagnetic exchange interaction between Fe and V is strong. However, it is well counteracted by the strong Fe-Fe interactions.    
In the MFA, the T$_C$ can be obtained as follows (not including factor $1/2$ in the Heisenberg Hamiltonian):
%-----------------------------
\begin{equation}
T_{C}^{\textrm{MFA}}=\frac{2J_0}{3k_B}.
\end{equation}
%------------------------------  
where $k_B$ is the Boltzmann constant. Here T$_C$ is proportional to the on-site exchange parameter J$_0$ where $J_0=\Sigma _iJ_{0i}$, i.e. the sum over all coordination shells. T$_C$s are rather similar in both compounds and well above the room temperature. Note that we only took into account Fe interactions for T$_C$ and possible contributions from Sm $4f$ are not included in the approximation since $4f$ are treated as spin-polarized core.

%------------------------------
%------------------------------ 

\subsection{Extrinsic magnetic properties}

\subsubsection{Stoner-Wohlfarth model}

\begin{figure}[h!]
\centering
\includegraphics[width=\columnwidth ,angle=0]{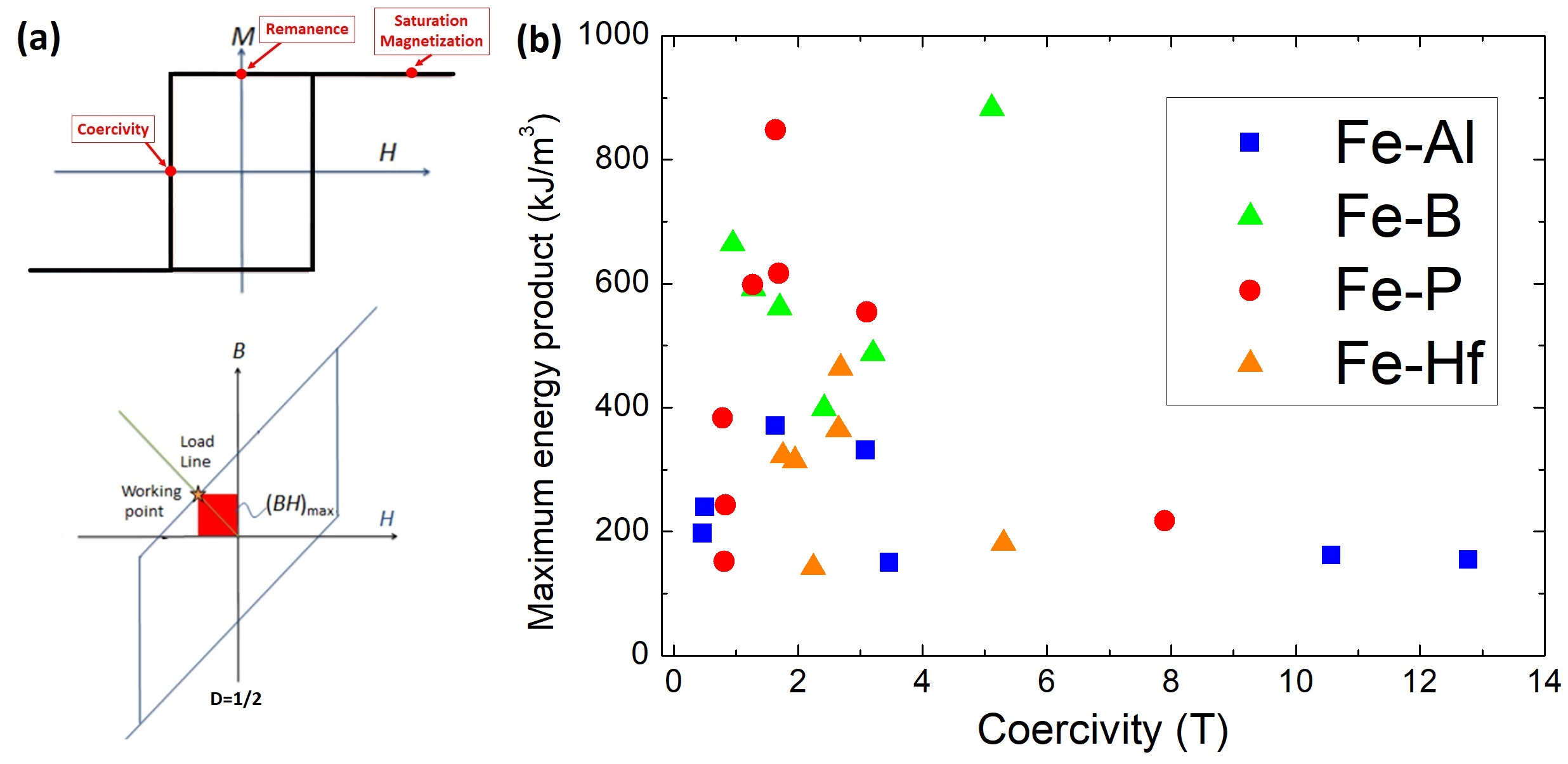}
\caption{(a) (top) Perfectly rectangular M(H) loop and (bottom) B(H) loop for obtaining maximum energy product. (b) High-throughput calculation of theoretical maximum energy product (BH)$_{max}$ and maximum coercivity ($H_{C,max}=H_K$) given by the Stoner-Wohlfarth model for the Fe-based binaries showed in Fig. \ref{fig:mae_flow} (with $\kappa>1/2$).}
\label{fig:stoner}
\end{figure}

Once the saturation magnetization ($M_S$) and magnetocrystalline anisotropy constants ($K_1$ and $K_2$) are calculated, it is possible to roughly estimate hysteresis loop properties as coercivity, remanence and maximum energy product by means of the Stoner-Wohlfarth model in the limit of coherent rotation \cite{Skomskibook}
%-----------------------------
\begin{equation}
\frac{E}{V}=K_1  \sin^2⁡\theta+K_2  \sin^4⁡\theta+\frac{\mu_0}{4} (1-3D) M_S^2  \sin^2⁡\theta-\mu_0 M_S H \cos\theta,
\end{equation}
%------------------------------      
where $E$ is the energy, $V$ is the volume of the material, $\theta$ is the angle between magnetization and the easy axis (along z-axis), $\mu_0$ is the vacuum permeability, $D$ is demagnetization factor (which can be a value between 0 and 1) and $H$ is the external magnetic field applied along the easy axis. The first two terms describe the magnetocrystalline energy (discussed in Section \ref{section:mae}), the third term is the shape anisotropy created by the demagnetizing field of the magnet and the last term is the Zeeman energy. In this model the remanence equals to saturation magnetization (perfectly rectangular hysteresis loop). Here, we define the anisotropy field as the minimum field needed to overcome the magnetocrytalline energy barrier (to reverse magnetization in the absence of shape anisotropy)
%-----------------------------
\begin{equation}
H_K=\frac{2(K_1+K_2)}{\mu_0 M_S} ,
\end{equation}
%------------------------------  
Note that there are other possible definitions \cite{Skomskibook}. Including the shape anisotropy, the minimum field to reverse the magnetization is the coercivity 
%-----------------------------
\begin{equation}
H_C=\frac{2(K_1+K_2)}{\mu_0 M_S}+\frac{1}{2} (1-3D) M_S^2. 
\end{equation}
%------------------------------  
This model describes well nano-size magnetic materials. Taking into account nucleation processes, one arrives to Brown relation
%-----------------------------
\begin{equation}
H_C\geq\frac{2K_1}{\mu_0 M_S}-DM_S.     
\end{equation}
%------------------------------  
The value of coercivity corresponding to uniform magnetization reversal by coherent rotation; generally, it is much lower than the anisotropy field, at best about 25\% of H$_K$. On the other hand, the energy product reads \cite{Coey2016} $(BH)=\mu_0 D(1-D)M_S^2$, which is maximum for $D=1/2$, that is,
%-----------------------------
\begin{equation} 
(BH)_{max}=\frac{1}{4}\mu_0 M_S^2.  
\end{equation}
%------------------------------  
Note this value can only be reached if $H_C>M_S/2$ (hardness parameter $\kappa>1/2$) \cite{Coey2016}. The database is being updated with values of magnetic extrinsic properties (remanence, coercivity and maximum energy product) given by the Stoner-Wohlfarth model (see Fig. \ref{fig:stoner}) for structures with easy axis, obtained in the high-throughput calculation of MAE (Section \ref{section:mae_AGA}).

%------------------------------
%------------------------------ 

\subsubsection{Micromagnetics}
\label{subsection:micromag}

In real magnets, to approach theoretical maximum energy product and coercivity it is essential to have an optimized microstructure. The  intergranular structure between the grains plays a significant role determining the magnetic properties, especially if the grain diameter is in the nanometer scale. Hard magnetic phases may be distorted locally near grain boundaries (GBs), because of the different lattice constant of the grain boundary phase. The distorted lattice may give rise to locally different intrinsic magnetic properties. Moreover, surface defects located at GB are sources of strong local demagnetizing fields, which act as nucleation centres. On the other hand, defects close to the GB can pin domain walls during the magnetization's reversal process. Nucleation and pinning mechanisms strongly affect extrinsic properties of hard magnetic phases, as the coercivity \cite{Kron}. Additionally, intergranular phases modify the exchange coupling behavior between the hard magnetic grains. For instance, nonmagnetic phases eliminate the direct exchange interaction between the hard magnetic grains. Typically, it leads to an increase of the coercive field. 
%------------------------------
\begin{figure}[h!]
\centering
\includegraphics[width=\columnwidth ,angle=0]{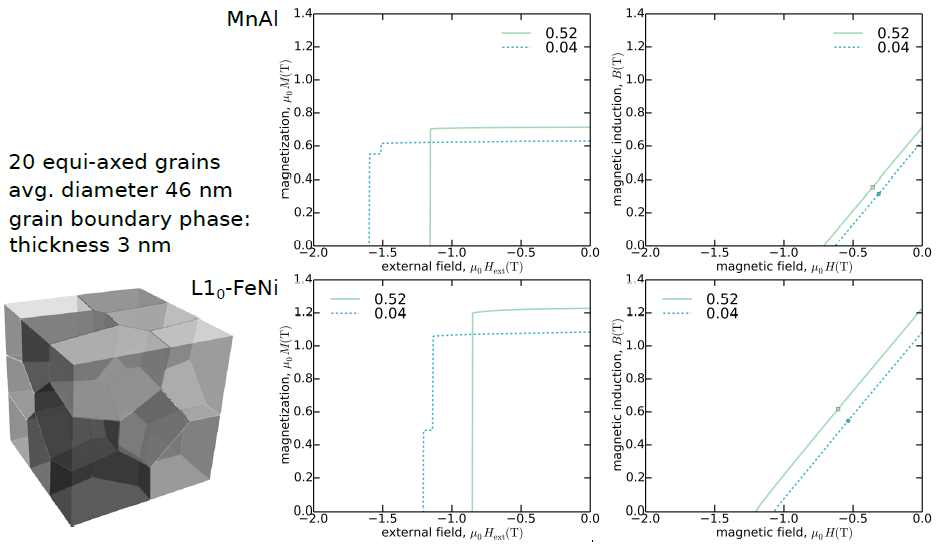}
\caption{(Left) Magnets structure with 3 nm thin grain boundary phase. (Mid) For both materials demagnetization curves M over H$_{ext}$ (Right) the corresponding B over H curve.}
\label{fig:micromag}
\end{figure}
%------------------------------ 

Novamag database contains theoretical results of extrinsic properties calculated with micromagnetic simulations considering optimized microstructures, constructed with the open-source 3D polycrystal generator tool Neper \cite{Neper}, for some of the magnets studied in Novamag project as Fe$_3$Sn$_{0.75}$Sb$_{0.25}$, L1$_0$ FeNi, MnAl $\tau$-phase, etc \cite{Kovacs}. In Fig. \ref{fig:micromag} we present demagnetization curves of such MnAl and L1$_0$ FeNi magnets highlighting the influence of different grain boundary magnetizations. As green solid line we show the results of M$_{S,gb}$ = 0.52$\cdot$M$_S$ and in blue dotted M$_{S,gb}$ = 0.04$\cdot$M$_S$ where M$_S$ is the main phase's saturation magnetization. The structure used for simulating the magnets consists of 20 equi-axed grains with an average grain size of 46 nm separated by a 3 nm thin layer denoted as the GB phase. With increased magnetization inside the GB phase the coercivity drops by 22\% and 16\% for MnAl and L1$_0$ FeNi respectively. In Fig. \ref{fig:micromag} on the right hand side we show the desheared BH loop using a shape factor of 1/3. Computed energy density products for L1$_0$ FeNi range from 230 kJ/m$^3$ to 300 kJ/m$^3$ whereas for MnAl ranges from 80 kJ/m$^3$ to 100 kJ/m$^3$ for high and low saturation magnetization inside the GB, respectively.

\subsection{Synthesis and characterization of hard magnetic phase alloys}

The comptutational high-throughput techniques described in previous sections can only approximately estimate the final properties and perfomance of a real bulk PM. On the other hand, there are also several high-throughput  experimental screening techniques as reactive crucible \cite{Fayyazi} or thin film combinatorial synthesis \cite{Gian}. However, they allow for a limited number of characterizations and are not representatives of the material in bulk, i.e. of the magnet. Therefore, whether the goal is the manufacture of a magnet, it is necessary to give the processing recipes of the material in bulk for a possible upscaling at an industrial level. 
%%%%%%%%%%%%%%%%%%%%%%%%%%%%%%%%%%%%%%%%%%
\begin{table}[]
  \centering
  \begin{adjustbox}{max width=\textwidth}
\begin{tabular}{|l|l|l|l|l|}
\hline
Sample                                       & Space Group & a(\text{\AA})  & b(\text{\AA})  & c(\text{\AA})  \\ \hline
Fe$_3$Sn                                     & $\#$194, P6$_3$/mmc         & 5.4621(5) & 5.4621(5) & 4.3490(6) \\ \hline
Fe$_{2.25}$Mn$_{0.75}$Sn$_{0.75}$Sb$_{0.25}$ & $\#$194, P6$_3$/mmc        & 5.4858(7) & 5.4858(7) & 4.3721(6) \\ \hline
Fe$_{2.5}$Mn$_{0.5}$Sn$_{0.75}$Sb$_{0.25}$   & $\#$194, P6$_3$/mmc         & 5.5551(1) & 5.5551(1) & 4.4398(1) \\ \hline
Fe$_2$Mn$_1$Sn$_{0.75}$Sb$_{0.25}$           & $\#$194, P6$_3$/mmc        & 5.5000(4) & 5.5000(4) & 4.3829(6) \\ \hline
Fe$_{1.5}$Mn$_{1.5}$Sn$_{0.75}$Sb$_{0.25}$   & $\#$194, P6$_3$/mmc        & 5.5338(1) & 5.5338(1) & 4.4270(2) \\ \hline
Fe$_{1.5}$Mn$_{1.5}$Sn$_{0.9}$Sb$_{0.1}$     & $\#$194, P6$_3$/mmc         & 5.5545(3) & 5.5545(3) & 4.4453(4) \\ \hline
\end{tabular}
\end{adjustbox}
\caption{Lattice parameters of the Fe$_{3-x}$Mn$_x$Sn$_{1-y}$Sb$_y$ alloys, obtained by Rietveld refinements of the X-Ray Diffraction patterns.}
  \label{tab:exp_Fe3Sn}
\end{table}
%%%%%%%%%%%%%%%%%%%%%%%%%%%%%%%%%%%%%%%%%%%%%

\begin{figure}[h!]
\centering
\includegraphics[width=\columnwidth ,angle=0]{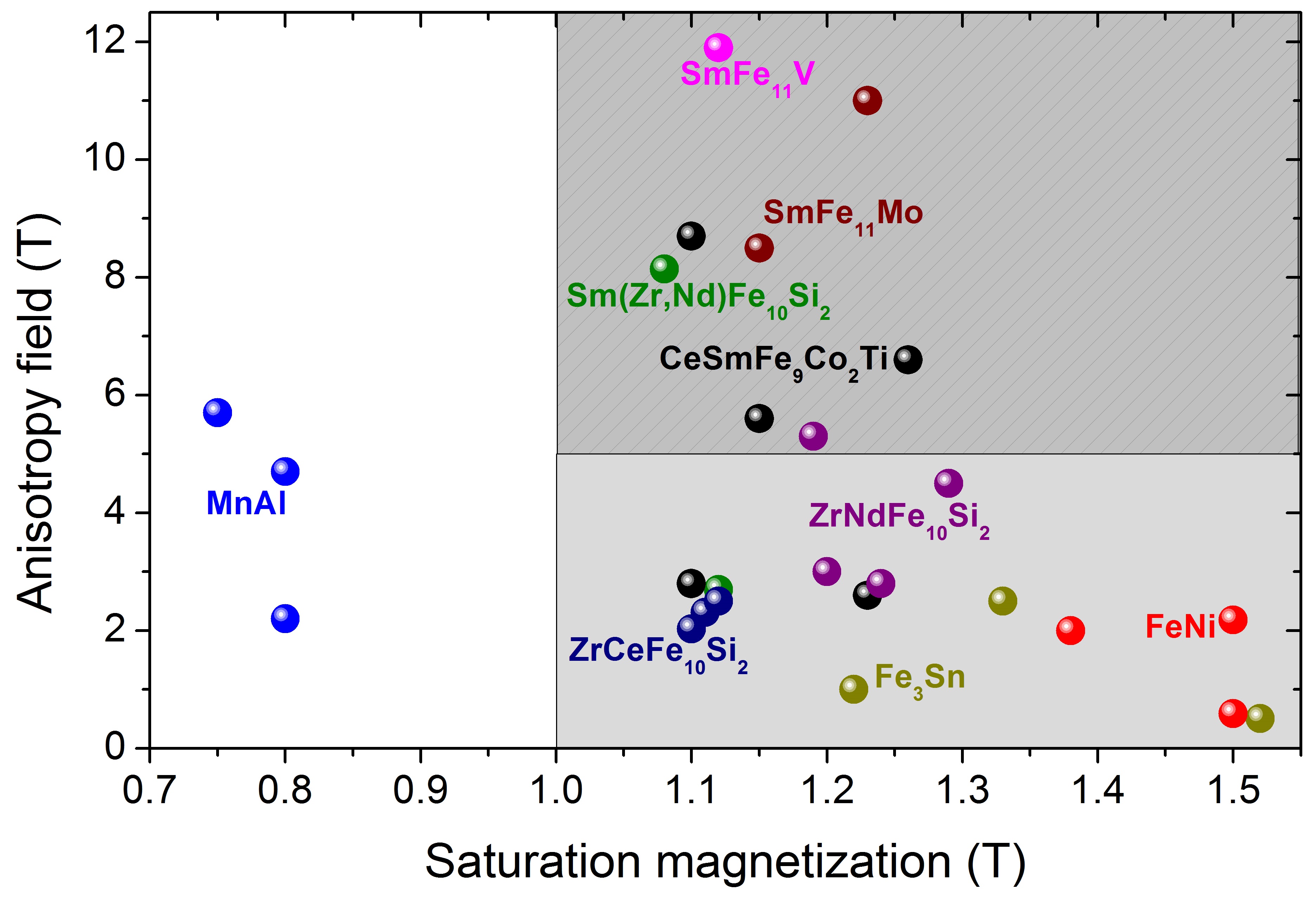}
\caption{Measured values of anisotropy field and saturation magnetization for some experimental magnetic samples included in the Novamag database.}
\label{fig:exp_Ms_Ha}
\end{figure}

In order to provide this kind of valuable information, the database is being extended with experimental results obtained in Novamag project, such as synthesis routes and the structural and magnetic characterization for CRM-free Fe$_3$Sn, FeNi and MnAl alloys \cite{Popov}, or the RE-lean ThMn$_{12}$-type tetragonal structure magnetic materials. For instance, Table \ref{tab:exp_Fe3Sn} shows the lattice parameters obtained by Rietveld refinements of the X-Ray Diffraction patterns for a set of Fe$_{3-x}$Mn$_x$Sn$_{1-y}$Sb$_y$ alloys synthesized by solid-state reaction  in the attempt to tune Fe$_3$Sn alloys \cite{Olga,Echevarria}, while Fig. \ref{fig:exp_Ms_Ha} shows a statistical summary of the anisotropy field and M$_S$ obtained in different alloys \cite{Martin1,Martin2,Gabay,Salazar,Madugundo}. We see that the alloys can be classified by their magnetic properties. As it is well known, the intrinsic properties of the magnetic alloys suitable for magnets must have high anisotropy field and high M$_S$ (dark grey lined zone) as SmFe$_{11}$V, SmFe$_{11}$Mo or SmFe$_{9}$Co$_2$Ti. Other alloys, which have high M$_S$, can be further processed in order to increase their anisotropy (light grey zone), as it is the case of Fe$_3$Sn. 

%------------------------------
%------------------------------ 

\section{Possible strategies to exploit promising theoretical novel phases in the Novamag database}

\label{section:strategy}

\begin{figure}[h!]
\centering
\includegraphics[width=\columnwidth ,angle=0]{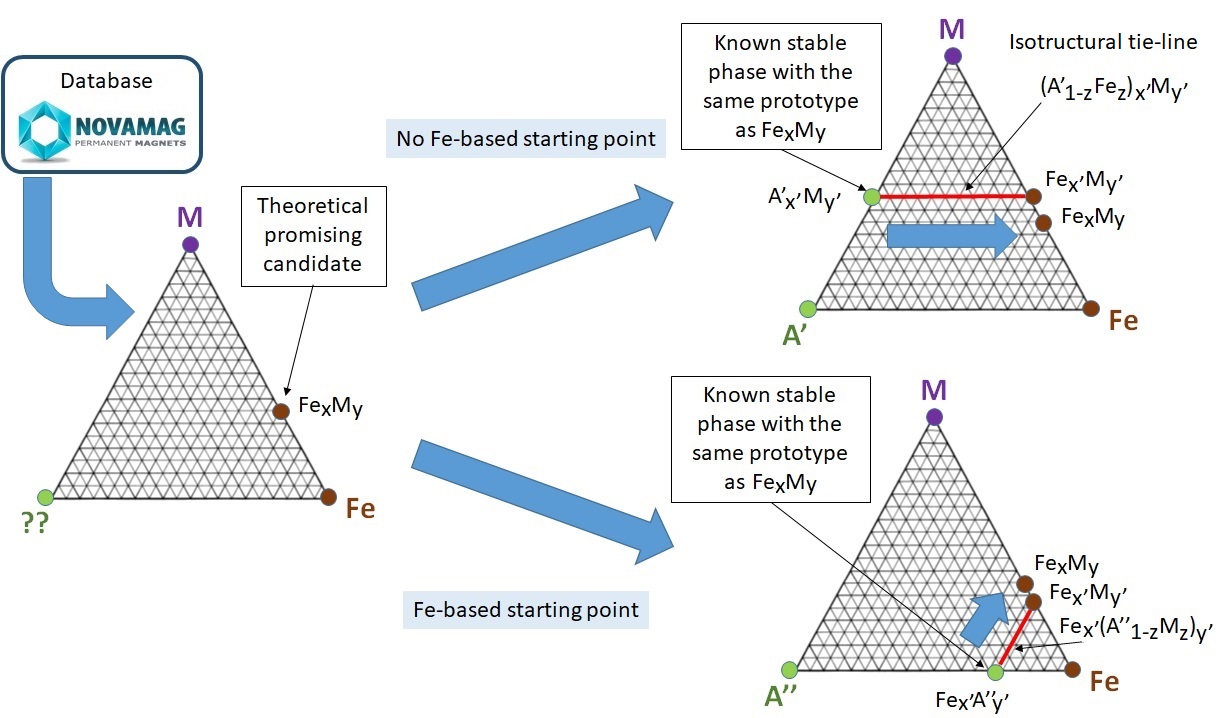}
\caption{Proposed general strategy to experimentally explore the crystal phase space neighborhood of promising theoretical Fe-based binaries found in the database.}
\label{fig:AGA_exp}
\end{figure}

MGI has created a new scenario with many theoretical phases that could be very appealing for different technological applications. However, one of the main challenges in MGI is to find experimental routes to synthesize most promising theoretical materials given by computational high-throughput  approaches. From a theoretical point of view, determining the stability of a predicted phase is a hard task that requires an accurate calculation of the full phase diagram at finite temperature (free energy) \cite{Liza,Zhao}. In this section, inspired by isostructural tie-lines \cite{Iga,Spriggs}, we discuss about some general strategies to design experimental procedures for exploring most promising theoretical materials found in the Novamag database, fully based on their crystallographic data, by doping known stable phases. The main idea is to think about the predicted phases by AGA in a more flexible manner, it means, as a hint or trend. Hence, making a magnet with some crystal and atomic similarities to these structures might be enough to get a good PM. Bearing this in mind, let's consider a given unknown Fe-based binary Fe$_x$M$_y$ generated by AGA that fulfills all necessary criteria for being a high-performance PM. First step is to search in literature (Handbook of crystallographic data or databases like Materials Project) for a known experimental stable binary phase that contains M element but no Fe (no Fe-based binary) A'$_{x'}$M$_{y'}$ or contains Fe but no M (Fe-based) Fe$_{x'}$A''$_{y'}$ with the same crystallographic prototype structure as the theoretical Fe$_x$M$_y$ and also similar stoichiometry. Next, these known stable phases (A'$_{x'}$M$_{y'}$ or Fe$_{x'}$A''$_{y'}$) are used as starting point to explore either the ternary line (A'$_{1-z}$Fe$_z$)$_{x'}$M$_{y'}$ or Fe$_{x'}$(A''$_{1-z}$M$_z$)$_{y'}$, that is, doping with Fe or M, respectively. In this way, it is possible to experimentally explore and approach the neighborhood of promising theoretical phase Fe$_x$M$_y$, where the compounds in this region with the same crystal structure might have similar properties as Fe$_x$M$_y$. Additionally, since this procedure is based on the same prototype structure as Fe$_x$M$_y$ it could also trigger the formation of this phase locally during the synthesis. Note that the route using a no Fe-based starting point might be more challenging since one needs to dope the ternary with high content of Fe in order to make it ferromagnetic, especially if A$’$ is not Co or Ni. Fig. \ref{fig:AGA_exp} shows a diagram explaining the above described strategy. Alternately, the selection of optimal substrates for epitaxial growth of these novel theoretical phases could be a way to stabilize them too \cite{Ding}.

%------------------------------
%------------------------------ 

\begin{figure}[h!]
\centering
\includegraphics[width=\columnwidth ,angle=0]{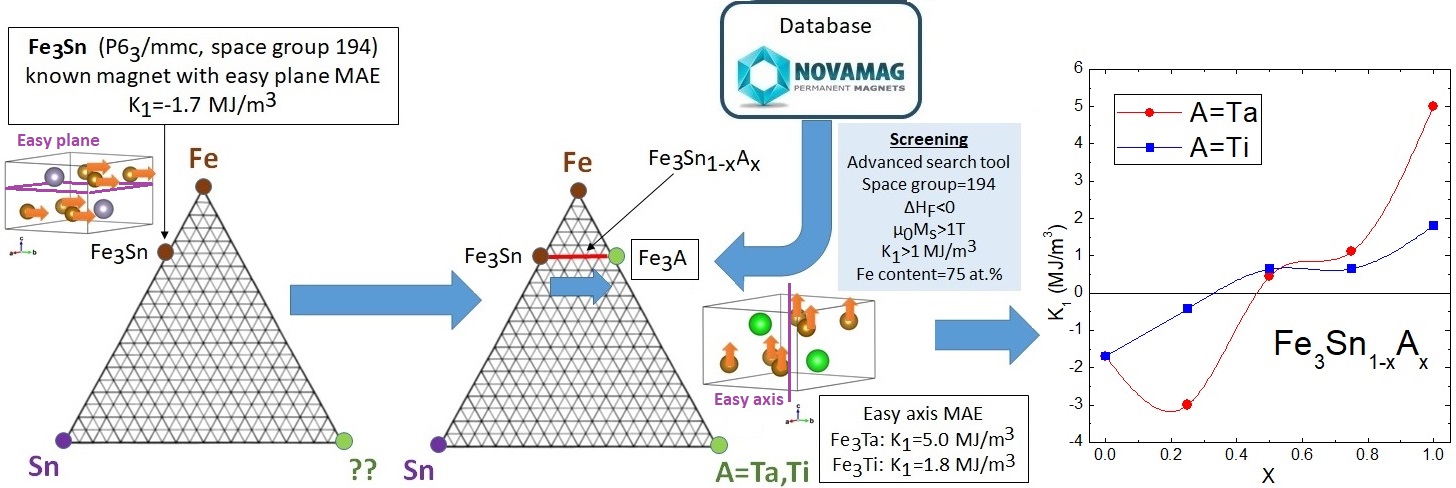}
\caption{Strategy to make use of Novamag database for tuning Fe$_3$Sn phase.}
\label{fig:AGA_exp_2}
\end{figure}

The above described strategy can be reversed. For instance, let's consider well-known phase Fe$_3$Sn (P6$_3$/mmc, space group 194) that exhibits an easy plane MAE. As indicated in Section \ref{subsection:UPP}, in this system it is necesseary to find a way to switch MAE from easy plane to easy axis in order to make it suitable for PM applications. To this end, we can use the search advanced tool of Novamag database\cite{novamag2} with the following filters: (i) space group=194, (ii) negative formation enthalpy $\Delta$H$_F<0$, (iii)  $\mu_0$M$_S>1$T, (iv) K$_1>$1 MJ/m$^3$, (v) binary alloy, and (vi) Fe content equal to 75 at.\%. Doing so, it found two promising phases Fe$_3$A with A=Ta,Ti that fulfill these requirements. Hence, it suggests to explore the ternary line Fe$_3$Sn$_{1-x}$A$_x$. A preliminary theoretical analysis using DFT calculations, following the steps described in Section \ref{section:mae_AGA}, with software VASP \cite{vasp_1,vasp_2,vasp_3} (PAW method \cite{PAW} and GGA-PBE exchange-correlation type \cite{PBE}) for 1x1x2 supercells (16 atoms) of Fe$_3$Sn$_{1-x}$A$_x$ (A=Ti,Ta) shows that replacing 50\% of Sn by Ti or Ta in the Fe$_3$Sn phase might be enough to switch MAE from easy plane to easy axis, see Fig. \ref{fig:AGA_exp_2}. Obviously, a more detailed analysis of the stability and MAE of these systems are required in order to clarify the role of these dopants on the Fe$3$Sn phase, see Section \ref{subsection:UPP} \cite{Olga}. 

%------------------------------
%------------------------------ 

Finally, concerning the cubic phases in which a tetragonal distortion can induce a large MAE (see Section \ref{subsection:tetragonal distortion}), we think that adding light interstitial atoms such as H, B, C, and N in these compounds might be a likely way to achieve it in practice \cite{Zhang:2016, Reichel:2017, Salikhov:2017}.

\section{Conclusions}

In summary, based on the philosophy and standards of MGI, we have developed a comprehensive open database of magnetic material parameters that is being updated with many significant theoretical and experimental results from Novamag project. In the design process, efforts have been made to use free software, standard vocabulary and metadata files that might facilitate interoperability with other existing databases. Additionally, its FOSS license maximizes openness and minimizes barriers to software use, dissemination, and follow-on innovation. 

Users worldwide can access to crystallographic and magnetic information of many theoretical structures calculated by DFT combined with an AGA. Presently, it is also being updated with the results of a high-throughput calculation of MAE for a large set of uniaxial magnetic structures predicted with AGA, as well as for tetragonally distorted cubic phases and tuned known magnetic materials. In order to illustrate these methods, we have reported some interesting phases with high MAE and M$_S$, such as FeAl, Fe$_{16}$B$_2$ and Fe$_4$Hf$_2$, tetragonally distorted Mn$_3$Pt and Mn$_2$RhPt, and doped Fe$_3$Sn.

Examples of available multiscale theoretical and experimental data about other relevant magnetic properties (exchange integrals, T$_C$, anisotropy field, coercivity, maximum energy product, etc.) were also shown, which can be helpful for the design of new promising high-performance RE-free/lean PMs. Aiming to link the computational high-throughput results to experiment, we presented some possible strategies to exploit most promising theoretical materials found in the database. Further results from the consortium of Novamag project are expected to be uploaded into the database in the near future, as well as possible new features and improvements.

\section*{Acknowledgements}

This work was supported by the European Horizon 2020 Framework Programme for Research and Innovation (2014-2020) under Grant Agreement No. 686056, NOVAMAG. The  authors  P. N. and S. A. thankfully  acknowledge the  computer  resources  at  Finisterrae and  the  technical  support  provided  by  Fundaci\'{o}n  P\'{u}blica  Galega  Centro  Tecnol\'{o}xico  de  Supercomputaci\'{o}n  de  Galicia (RES-QCM-2018-3-0009). O. Y. V., H. C. H. and O. E. thankfully  acknowledge the  computer resources provided by the Swedish National Infrastructure for Computing (SNIC) at PDC and NSC centers. O. E. acknowledges support from STandUP, eSSENCE, the Swedish Research Council and the Knut and Alice Wallenberg foundation (KAW) and the foundation for strategic research (SSF).

\bibliography{mybibfile}

\begin{thebibliography}{100}

\bibitem{Gutfleisch2011}
O.~Gutfleisch, M.~A. Willard, E.~Br{\"u}ck, C.~H. Chen, S.~G. Sankar, and
  J.~Ping Liu.
\newblock Magnetic materials and devices for the 21st century: Stronger,
  lighter, and more energy efficient.
\newblock {\em Adv. Mater.}, 23:821--842, 2011.

\bibitem{Coeybook}
J.~M.~D. Coey.
\newblock {\em Magnetism and Magnetic Materials}.
\newblock Cambridge University Press, 2010.

\bibitem{Lewis2013}
L.~H. Lewis and F.~Jim{\'e}nez-Villacorta.
\newblock Perspectives on permanent magnetic materials for energy conversion
  and power generation.
\newblock {\em Metall. Mater. Trans. A}, 44(Suppl 1):2, 2013.

\bibitem{Anja}
Anja Brumme.
\newblock {\em Wind Energy Deployment and the Relevance of Rare Earths}.
\newblock Springer, 2014.

\bibitem{Skomskibook}
R.~Skomski and J.~M.~D. Coey.
\newblock {\em Permanent magnetism}.
\newblock Institute of Physics Publishing, 1999.

\bibitem{Coey2016}
R.~Skomski and J.~M.~D. Coey.
\newblock Magnetic anisotropy --- {H}ow much is enough for a permanent magnet?
\newblock {\em Scripta Materialia}, 112:3--8, 2016.

\bibitem{Kron}
H.~Kronm{\"u}ller and M.~F{\"a}hnle.
\newblock {\em Micromagnetism and the Microstructure of Ferromagnetic Solids}.
\newblock Cambridge University Press, 2003.

\bibitem{Coey2014}
J.~M.~D. Coey.
\newblock New permanent magnets; manganese compounds.
\newblock {\em J. Phys.: Condens. Matter}, 26:064211, 2014.

\bibitem{McCallum}
R.W. McCallum, L.H. Lewis, R.~Skomsky, M.J. Kramer, and I.E. Anderson.
\newblock Practical aspects of modern and future permanent magnets.
\newblock {\em Annu. Rev. Mater. Res.}, 44:451, 2014.

\bibitem{Kuzmin:2014}
M~D Kuz'min, K~P Skokov, H~Jian, I~Radulov, and O~Gutfleisch.
\newblock Towards high-performance permanent magnets without rare earths.
\newblock {\em Journal of Physics: Condensed Matter}, 26(6):064205, jan 2014.

\bibitem{Skokov:2018}
K.P. Skokov and O.~Gutfleisch.
\newblock Heavy rare earth free, free rare earth and rare earth free magnets -
  vision and reality.
\newblock {\em Scripta Materialia}, 154:289 -- 294, 2018.

\bibitem{Drebov}
N.~Drebov, A.~Martinez-Limia, L.~Kunz, A.~Gola, T.~Shigematsu, T.~Eckl,
  P.~Gumbsch, and C.~Els{\"a}sser.
\newblock Ab initio screening methodology applied to the search for new
  permanent magnetic materials.
\newblock {\em New Journal of Physics}, 15:125023, 2013.

\bibitem{Korner}
W.~K{\"o}rner, G.~Krugel, and C.~Els{\"a}sser.
\newblock Theoretical screening of intermetallic {T}h{M}n$_{12}$-type phases
  for new hard-magnetic compounds with low rare earth content.
\newblock {\em Scientific Reports}, 6:24686, 2016.

\bibitem{MGI_1}
Materials genome initiative strategic plan. executive office of the president
  of the united states of america. december 2014.
\newblock \url{http://www.nist.gov/mgi/upload/MGI-StrategicPlan-2014.pdf}.

\bibitem{MGI_2}
X.~Qu and et~al.
\newblock Perspective on materials genome.
\newblock {\em Chin. Sci. Bull.}, 59:1619--1623, 2014.

\bibitem{Aflow_1}
Aflowlib.
\newblock http://aflowlib.org/.

\bibitem{Aflow_2}
S.~Curtarolo, W.~Setyawan, S.~Wang, J.~Xue, K.~Yang, R.~H. Taylor, L.~J.
  Nelson, G.~L.W. Hart, S.~Sanvito, M.~Buongiorno-Nardelli, N.~Mingo, and
  O.~Levy.
\newblock Aflowlib.org: A distributed materials properties repository from
  high-throughput ab initio calculations.
\newblock {\em Comp. Mater. Sci.}, 58:227, 2012.

\bibitem{Mat_Proj_1}
A.~Jain, S.P. Ong, G.~Hautier, W.~Chen, W.D. Richards, S.~Dacek, S.~Cholia,
  D.~Gunter, D.~Skinner, G.~Ceder, and K.A. Persson.
\newblock The materials project: A materials genome approach to accelerating
  materials innovation.
\newblock {\em APL Materials}, 1:011002, 2013.

\bibitem{Mat_Proj_2}
The materials project.
\newblock https://materialsproject.org/.

\bibitem{LU2017191}
Wencong Lu, Ruijuan Xiao, Jiong Yang, Hong Li, and Wenqing Zhang.
\newblock Data mining-aided materials discovery and optimization.
\newblock {\em Journal of Materiomics}, 3(3):191 -- 201, 2017.

\bibitem{LIU2017159}
Yue Liu, Tianlu Zhao, Wangwei Ju, and Siqi Shi.
\newblock Materials discovery and design using machine learning.
\newblock {\em Journal of Materiomics}, 3(3):159 -- 177, 2017.

\bibitem{nomad_coe}
\url{https://repository.nomad-coe.eu/}.

\bibitem{Jong_1}
Maarten de~Jong, Wei Chen, Thomas Angsten, Anubhav Jain, Randy Notestine,
  Anthony Gamst, Marcel Sluiter, Chaitanya Krishna~Ande, Sybrand van~der Zwaag,
  Jose~J Plata, Cormac Toher, Stefano Curtarolo, Gerbrand Ceder, Kristin~A.
  Persson, and Mark Asta.
\newblock Charting the complete elastic properties of inorganic crystalline
  compounds.
\newblock {\em Scientific Data}, 2, 03 2015.

\bibitem{Jong_2}
Maarten de~Jong, Wei Chen, Henry Geerlings, Mark Asta, and Kristin~Aslaug
  Persson.
\newblock A database to enable discovery and design of piezoelectric materials.
\newblock {\em Scientific Data}, 2, 09 2015.

\bibitem{Qu}
Xiaohui Qu, Anubhav Jain, Nav~Nidhi Rajput, Lei Cheng, Yong Zhang, Shyue~Ping
  Ong, Miriam Brafman, Edward Maginn, Larry~A. Curtiss, and Kristin~A. Persson.
\newblock The electrolyte genome project: A big data approach in battery
  materials discovery.
\newblock {\em Computational Materials Science}, 103:56 -- 67, 2015.

\bibitem{Rev_database}
K.~T. Butler and et~al.
\newblock Machine learning for molecular and materials science.
\newblock {\em Nature}, 559:549, 2018.

\bibitem{Magndata}
http://webbdcrista1.ehu.es/magndata/.

\bibitem{data_Tc}
C.J. Court and J.M. Cole.
\newblock Auto-generated materials database of {C}urie and {N}{\'e}el
  temperatures via semisupervised relationship extraction.
\newblock {\em Scientific Data}, 5:180111, 2018.

\bibitem{novamag_web}
\url{http://www.novamag.eu/}.

\bibitem{novamag}
\url{http://crono.ubu.es/novamag/}.

\bibitem{web}
\url{https://github.com/rmartico/NOVAMAG-UBU-WEB}.

\bibitem{loader}
\url{https://github.com/rmartico/NOVAMAG-Java-Loader}.

\bibitem{postgresql}
\url{https://www.postgresql.org/}.

\bibitem{fhi}
\url{https://th.fhi-berlin.mpg.de/site/uploads/Publications/Psik_Highlight_131-2016.pdf}.

\bibitem{tiobe}
\url{www.tiobe.com}.

\bibitem{pypi}
\url{https://pypi.org/project/sqlacodegen/}.

\bibitem{python}
\url{https://www.python.org/download/releases/3.0/}.

\bibitem{flask}
\url{http://flask.pocoo.org/}.

\bibitem{pythonhosted}
\url{https://pythonhosted.org/Flask-Bootstrap/}.

\bibitem{flask2}
\url{http://flask-sqlalchemy.pocoo.org/2.3/}.

\bibitem{jetbrains}
\url{https://www.jetbrains.com/pycharm/}.

\bibitem{ecma}
\url{http://www.ecma-international.org/publications/files/ECMA-ST/ECMA-404.pdf}.

\bibitem{oracle}
\url{http://www.oracle.com/technetwork/java/javase/overview/index.html}.

\bibitem{oracle2}
\url{https://docs.oracle.com/javase/8/docs/technotes/guides/jdbc/}.

\bibitem{oracle3}
\url{https://docs.oracle.com/javase/8/docs/technotes/guides/jndi/index.html}.

\bibitem{apache}
\url{https://logging.apache.org/log4j/2.x/}.

\bibitem{slf4j}
\url{https://www.slf4j.org/}.

\bibitem{junit}
\url{https://junit.org/junit4/}.

\bibitem{github}
\url{https://github.com/}.

\bibitem{eclipse}
\url{https://www.eclipse.org/mars/}.

\bibitem{novamag2}
\url{http://crono.ubu.es/novamag/advanced_search}.

\bibitem{Oganov}
A.~R. Oganov.
\newblock {\em Modern Methods of Crystal Structure Prediction}.
\newblock WILEY-VCH Verlag GmbH and Co. KGaA, Weinheim, 2011.

\bibitem{uspex}
Andriy~O. Lyakhov, Artem~R. Oganov, Harold~T. Stokes, and Qiang Zhu.
\newblock New developments in evolutionary structure prediction algorithm
  uspex.
\newblock {\em Computer Physics Communications}, 184(4):1172 -- 1182, 2013.

\bibitem{vasp_1}
G.~Kresse and J.~Hafner.
\newblock Ab initio molecular dynamics for liquid metals.
\newblock {\em Phys. Rev. B}, 47:558--561, Jan 1993.

\bibitem{vasp_2}
G.~Kresse and J.~Furthmüller.
\newblock Efficiency of ab-initio total energy calculations for metals and
  semiconductors using a plane-wave basis set.
\newblock {\em Computational Materials Science}, 6(1):15 -- 50, 1996.

\bibitem{vasp_3}
G.~Kresse and J.~Furthm\"uller.
\newblock Efficient iterative schemes for ab initio total-energy calculations
  using a plane-wave basis set.
\newblock {\em Phys. Rev. B}, 54:11169--11186, Oct 1996.

\bibitem{Arapan_AGA_1}
S.~Arapan, P.~Nieves, and S.~Cuesta-L{\'o}pez.
\newblock A high-throughput exploration of magnetic materials by using
  structure predicting methods.
\newblock {\em J. Appl. Phys.}, 123:083904, 2018.

\bibitem{Arapan_AGA_2}
P.~Nieves, S.~Arapan, G.~C. Hadjipanayis, D.~Niarchos, J.~M. Barandiaran, and
  S.~Cuesta-L{\'o}pez.
\newblock Applying high-throughput computational techniques for discovering
  next-generation of permanent magnets.
\newblock {\em Phys. Status Solidi C}, 13:942--950, 2016.

\bibitem{Arapan_AGA_3}
P.~Nieves, S.~Arapan, and S.~Cuesta-L{\'o}pez.
\newblock Exploring the crystal structure space of {C}o{F}e$_{2}${P} by using
  adaptive genetic algorithm methods.
\newblock {\em IEEE Transactions on Magnetics}, 53:11, 2017.

\bibitem{Burkert:2004}
Till Burkert, Lars Nordstr\"om, Olle Eriksson, and Olle Heinonen.
\newblock Giant magnetic anisotropy in tetragonal {F}e{C}o alloys.
\newblock {\em Phys. Rev. Lett.}, 93:027203, Jul 2004.

\bibitem{Zhang:2016}
Hongbin Zhang, Imants Dirba, Tim Helbig, Lambert Alff, and Oliver Gutfleisch.
\newblock Engineering perpendicular magnetic anisotropy in {F}e via
  interstitial nitrogenation: {N} choose {K}.
\newblock {\em APL Materials}, 4(11):116104, 2016.

\bibitem{Reichel:2017}
L~Reichel, A~Edstr{\"o}m, D~Pohl, J~Rusz, O~Eriksson, L~Schultz, and
  S~F{\"a}hler.
\newblock On the origin of perpendicular magnetic anisotropy in strained
  {F}e{\textendash}{C}o({\textendash}{X}) films.
\newblock {\em Journal of Physics D: Applied Physics}, 50(4):045003, jan 2017.

\bibitem{Salikhov:2017}
R~Salikhov, L~Reichel, B~Zingsem, R~Abrudan, A~Edstr{\"o}m, D~Thonig, J~Rusz,
  O~Eriksson, L~Schultz, S~F{\"a}hler, M~Farle, and U~Wiedwald.
\newblock Enhanced spin{\textendash}orbit coupling in tetragonally strained
  {F}e{\textendash}{C}o{\textendash}{B} films.
\newblock {\em Journal of Physics: Condensed Matter}, 29(27):275802, jun 2017.

\bibitem{Olga}
O.~Yu. Vekilova, B.~Fayyazi, K.~P. Skokov, O.~Gutfleisch, C.~Echevarria-Bonet,
  J.~M. Barandiar{\'a}n, A.~Kovacs, J.~Fischbacher, T.~Schrefl, O.~Eriksson,
  and H.~C. Herper.
\newblock Tuning magnetocrystalline anisotropy of {F}e$_{3}${S}n by alloying.
\newblock {\em Phys. Rev. B}, 99:024421, 2019.

\bibitem{Sales}
Brian~C. Sales, Bayrammurad Saparov, Michael~A. McGuire, David~J. Singh, and
  David~S. Parker.
\newblock Ferromagnetism of {F}e$_{3}${S}n and {A}lloys.
\newblock {\em Scientific Reports}, 4:7024 EP --, 11 2014.

\bibitem{PAW}
P.~E. Bl\"ochl.
\newblock Projector augmented-wave method.
\newblock {\em Phys. Rev. B}, 50:17953--17979, Dec 1994.

\bibitem{PBE}
John~P. Perdew, Kieron Burke, and Matthias Ernzerhof.
\newblock Generalized gradient approximation made simple.
\newblock {\em Phys. Rev. Lett.}, 77:3865--3868, Oct 1996.

\bibitem{RSPT_1}
John~M. Wills and Bernard~R. Cooper.
\newblock Synthesis of band and model {H}amiltonian theory for hybridizing
  cerium systems.
\newblock {\em Phys. Rev. B}, 36:3809--3823, Sep 1987.

\bibitem{RSPt_2}
J.~M. Wills, M.~Alouani, P.~Andersson, A.~Delin, O.~Eriksson, and O.~Grechnyev.
\newblock {\em Full-{P}otential {E}lectronic {S}tructure {M}ethod}, volume 167
  of {\em Springer series in solid state science}.
\newblock Springer, Berlin, Germany, 2010.

\bibitem{bloechl_tetra}
Peter~E. Bl\"ochl, O.~Jepsen, and O.~K. Andersen.
\newblock Improved tetrahedron method for {B}rillouin-zone integrations.
\newblock {\em Phys. Rev. B}, 49:16223--16233, Jun 1994.

\bibitem{monkhorst-pack}
Hendrik~J. Monkhorst and James~D. Pack.
\newblock Special points for {B}rillouin-zone integrations.
\newblock {\em Phys. Rev. B}, 13:5188--5192, Jun 1976.

\bibitem{Eriksson_book}
O.~Eriksson, A.~Bergman, L.~Bergqvist, and J.~Hellsvik.
\newblock {\em Atomistic Spin Dynamics Foundations and Applications}.
\newblock Oxford University Press, 2017.

\bibitem{Kurz}
Ph. Kurz, F.~F{\"o}rster, L.~Nordstr{\"o}m, G.~Bihlmayer, and S.~Bl{\"u}gel.
\newblock Ab initio treatment of noncollinear magnets with the full-potential
  linearized augmented plane wave method.
\newblock {\em Phys. Rev. B}, 69:024415, 2004.

\bibitem{Skubic}
B~Skubic, J~Hellsvik, L~Nordström, and O~Eriksson.
\newblock A method for atomistic spin dynamics simulations: implementation and
  examples.
\newblock {\em Journal of Physics: Condensed Matter}, 20(31):315203, jul 2008.

\bibitem{uppasd}
\url{http://www.physics.uu.se/research/materials-theory/ongoingresearch/uppasd/}.

\bibitem{Evans}
R.~F.~L. Evans, W.~J. Fan, P.~Chureemart, T.~A. Ostler, M.~O.~A. Ellis, and
  R.W. Chantrell.
\newblock Atomistic spin model simulations of magnetic nanomaterials.
\newblock {\em J. Phys.: Condens.Matter}, 26:103202, 2014.

\bibitem{vampire}
\url{https://vampire.york.ac.uk/}.

\bibitem{Nieves}
P.~Nieves, S.~Arapan, T.~Schrefl, and S.~Cuesta-L{\'o}pez.
\newblock Atomistic spin dynamics simulations of the {M}n{A}l ${\tau}$-phase
  and its antiphase boundary.
\newblock {\em Phys. Rev. B}, 96:224411, 2017.

\bibitem{Jacob}
A.~Jakobsson, P.~Mavropoulos, E.~\ifmmode \mbox{\c{S}}\else \c{S}\fi{}a\ifmmode
  \mbox{\c{s}}\else \c{s}\fi{}\ifmmode \imath \else \i
  \fi{}o\ifmmode~\breve{g}\else \u{g}\fi{}lu, S.~Bl\"ugel, M.~Le\ifmmode
  \check{z}\else \v{z}\fi{}ai\ifmmode~\acute{c}\else \'{c}\fi{}, B.~Sanyal, and
  I.~Galanakis.
\newblock First-principles calculations of exchange interactions, spin waves,
  and temperature dependence of magnetization in inverse-{H}eusler-based spin
  gapless semiconductors.
\newblock {\em Phys. Rev. B}, 91:174439, May 2015.

\bibitem{Lezaic}
M.~Le{\v{z}}ai{\'c}.
\newblock {\em Spin-gap Materials from First Principles: Properties and
  Applications of Half-metallic Ferromagnets}.
\newblock PhD thesis, Rheinisch-Westf{\"a}lischen Technischen Hochschule
  Aachen, 2005.

\bibitem{Lewis}
Laura~H. Lewis, Frederick~E. Pinkerton, Nina Bordeaux, Arif Mubarok, Eric
  Poirier, Joseph~I. Goldstein, Ralph Skomski, and Katayun Barmak.
\newblock De magnete et meteorite: Cosmically motivated materials.
\newblock {\em IEEE MAGNETICS LETTERS}, 5:5500104, 2014.

\bibitem{Evans2}
R.~F.~L. Evans, U.~Atxitia, and R.~W. Chantrel.
\newblock Quantitative simulation of temperature-dependent magnetization
  dynamics and equilibrium properties of elemental ferromagnets.
\newblock {\em Phys. Rev. B}, 91:144425, 2015.

\bibitem{Woo}
C.~H. Woo, Haohua Wen, A.~A. Semenov, S.~L. Dudarev, and Pui-Wai Ma.
\newblock Quantum heat bath for spin-lattice dynamics.
\newblock {\em Phys. Rev. B}, 91:104306, 2015.

\bibitem{Kormann}
F.~K{\"o}rmann, A.~Dick, T.~Hickel, and J.~Neugebauer.
\newblock Role of spin quantization in determining the thermodynamic properties
  of magnetic transition metals.
\newblock {\em Phys. Rev. B}, 83:165114, 2011.

\bibitem{Kazan}
N.~Kazantseva, D.~Hinzke, U.~Nowak, R.~W. Chantrell, U.~Atxitia, and
  O.~Chubykalo-Fesenko.
\newblock Towards multiscale modeling of magnetic materials: Simulations of
  {F}e{P}t.
\newblock {\em Phys. Rev. B}, 77:184428, 2008.

\bibitem{Ana_Maria}
A.M. Sch{\"o}nh{\"o}bel, R.~Madugundo, O.~Yu. Vekilova, O.~Eriksson, H.C.
  Herper, J.M. Barandiarán, and G.C. Hadjipanayis.
\newblock Intrinsic magnetic properties of {S}m{F}e${12−x}${V}$_{x}$ alloys
  with reduced {V}-concentration.
\newblock {\em Journal of Alloys and Compounds}, 786:969 -- 974, 2019.

\bibitem{Jij_1}
M.~I. Katsnelson and A.~I. Lichtenstein.
\newblock First-principles calculations of magnetic interactions in correlated
  systems.
\newblock {\em Phys. Rev. B}, 61:8906--8912, Apr 2000.

\bibitem{Jij_2}
A.I. Liechtenstein, M.I. Katsnelson, V.P. Antropov, and V.A. Gubanov.
\newblock Local spin density functional approach to the theory of exchange
  interactions in ferromagnetic metals and alloys.
\newblock {\em Journal of Magnetism and Magnetic Materials}, 67(1):65 -- 74,
  1987.

\bibitem{Jij_3}
Y.~O. Kvashnin, O.~Gr\aa{}n\"as, I.~Di~Marco, M.~I. Katsnelson, A.~I.
  Lichtenstein, and O.~Eriksson.
\newblock Exchange parameters of strongly correlated materials: Extraction from
  spin-polarized density functional theory plus dynamical mean-field theory.
\newblock {\em Phys. Rev. B}, 91:125133, Mar 2015.

\bibitem{Neper}
R.~Quey, P.R. Dawson, and F.~Barbe.
\newblock Large-scale 3d random polycrystals for the finite element method:
  {G}eneration, meshing and remeshing.
\newblock {\em Computer Methods in Applied Mechanics and Engineering},
  200:1729--1745, 2011.

\bibitem{Kovacs}
A.~Kovacs, J.~Fischbacher, M.~Gusenbauer, H.~Oezelt, H.~C. Herper, O.~Y.
  Vekilova, P.~Nieves, S.~Arapan, and T.~Schrefl.
\newblock Computational {D}esign of the {R}are-{E}arth {R}educed {P}ermanent
  {M}agnets.
\newblock {\em Engineering, proceedings of the REMP2018 conference}, 2018.

\bibitem{Fayyazi}
B.~Fayyazi, K.P. Skokov, T.~Faske, D.Y. Karpenkov, W.~Donner, and
  O.~Gutfleisch.
\newblock Bulk combinatorial analysis for searching new rare-earth free
  permanent magnets: {R}eactive crucible melting applied to the {F}e-{S}n
  binary system.
\newblock {\em Acta Mater.}, 141:434–--443, 2017.

\bibitem{Gian}
G.~Giannopoulos, G.~Barucca, A.~Kaidatzis, V.~Psycharis, R.~Salikhov, M.~Farle,
  E.~Koutsouflakis, D.~Niarchos, A.~Mehta, M.~Scuderi, G.~Nicotra, C.~Spinella,
  S.~Laureti, and G.~Varvaro.
\newblock L1$_{0}$-{F}e{N}i films on {A}u-{C}u-{N}i buffer-layer: a
  high-throughput combinatorial study.
\newblock {\em Sci. Rep.}, 8:15919, 2018.

\bibitem{Popov}
V.~Popov, A.~Koptyug, I.~Radulov, F.~Maccari, and G.~Muller.
\newblock Prospects of additive manufacturing of rare-earth and non-rare-earth
  permanent magnets.
\newblock {\em Procedia Manufacturing}, 21:100--108, 2018.

\bibitem{Echevarria}
C.~Echevarria-Bonet, N.~Iglesias, J.S. Garitaonandia, D.~Salazar, G.C.
  Hadjipanayis, and J.M. Barandiaran.
\newblock Structural and magnetic properties of hexagonal {F}e$_{3}${S}n
  prepared by non-equilibrium techniques.
\newblock {\em Journal of Alloys and Compounds}, 769:843--847, 2018.

\bibitem{Martin1}
A.~Martin-Cid, A.M. Gabay, D.~Salazar, J.M. Barandiaran, and G.C. Hadjipanayis.
\newblock Tetragonal {C}e-based {C}e-{S}m({F}e,{C}o,{T}i)$_{12}$ alloys for
  permanent magnets.
\newblock {\em Phys. Status Solidi Curr. Top. Solid State Phys.}, page 15919,
  2016.

\bibitem{Martin2}
A.~Martin-Cid, D.~Salazar, A.M. Sch{\"o}nh{\"o}bel, J.S. Garitaonandia, J.M.
  Barandiaran, and G.C. Hadjipanayis.
\newblock Magnetic properties and phase stability of tetragonal
  {C}e$_{1-x}${S}m$_{x}${F}e$_{9}${C}o$_{2}${T}i 1:12 phase for permanent
  magnets.
\newblock {\em J. Alloys Compd.}, 749:640–--644, 2018.

\bibitem{Gabay}
A.M. Gabay, A.~Martín-Cid, J.M. Barandiaran, D.~Salazar, and G.C.
  Hadjipanayis.
\newblock Low-cost {C}e$_{1-x}${S}m$_{x}$({F}e,{C}o,{T}i)$_{12}$ alloys for
  permanent magnets.
\newblock {\em AIP Adv.}, 6:056015, 2016.

\bibitem{Salazar}
D.~Salazar, A.~Martín-Cid, J.S. Garitaonandia, T.C. Hansen, J.M. Barandiaran,
  and G.C. Hadjipanayis.
\newblock Role of {C}e substitution in the magneto-crystalline anisotropy of
  tetragonal {Z}r{F}e$_{10}${S}i$_{2}$.
\newblock {\em J. Alloys Compd.}, 766:291–--296, 2018.

\bibitem{Madugundo}
R.~Madugundo, N.V.~Rama Rao, A.M. Sch{\"o}nh{\"o}bel, D.~Salazar, and A.A.
  El-Gendy.
\newblock Recent developments in nanostructured permanent magnet materials and
  their processing methods.
\newblock {\em Magn. Nanostructured Mater.}, pages 157--198, 2018.

\bibitem{Liza}
R.~Liz{\'a}rraga, F.~Pan, L.~Bergqvist, E.~Holmstr{\"o}m, Z.~Gercsi, and
  L.~Vitos.
\newblock First principles theory of the hcp-fcc phase transition in {C}obalt.
\newblock {\em Scientific Reports}, 7:3778, 2017.

\bibitem{Zhao}
Xin Zhao, Shu Yu, Shunqing Wu, Manh~Cuong Nguyen, Cai-Zhuang Wang, and Kai-Ming
  Ho.
\newblock Structures, phase transitions, and magnetic properties of
  {C}o$_{3}${S}i from first-principles calculations.
\newblock {\em Phys. Rev. B}, 96:024422, 2017.

\bibitem{Iga}
A.~Iga.
\newblock Magnetocrystalline anisotropy in ({F}e$_{1-x}${C}o$_{x}$)$_{2}${B}
  system.
\newblock {\em Japan {J}. {A}ppl. {P}hys.}, 9:415--416, 1970.

\bibitem{Spriggs}
P.~H. Spriggs.
\newblock An investigation of the variation of lattice parameters with
  composition along the tie-line {N}i$_{3}${P}-{F}e$_{3}${P}.
\newblock {\em Philosophical Magazine}, 21:173, 1970.

\bibitem{Ding}
H.~Ding, S.~S. Dwaraknath, L.~Garten, P.~Ndione, D.~Ginley, and K.~A. Persson.
\newblock Computational approach for epitaxial polymorph stabilization through
  substrate selection.
\newblock {\em ACS Applied Materials $\&$ Interfaces}, 8(20):13086--13093,
  2016.

\end{thebibliography}

\end{document}